\documentclass[%
reprint,
showpacs,preprintnumbers,
 amsmath,amssymb,
 aps,
pra,
]{revtex4-1}
\usepackage{braket}
\usepackage[usenames, dvipsnames]{color}
\usepackage{graphicx}
\usepackage{dcolumn}
\usepackage{bm}
\usepackage{hyperref}

\usepackage{caption}

\begin{document}


\title{Population dynamics in sideband cooling of trapped ions outside the Lamb-Dicke regime}

\author{M. K. Joshi} 
\altaffiliation[Also at:-]{Institute for Quantum Optics Quantum Information, Innsbruck, Austria}
\author{ P. Hrmo} \altaffiliation[Also at:-]{University of Innsbruck, Innsbruck, Austria}
\author{ V. Jarlaud}
\altaffiliation[Also at:-]{Aarhus University,
Aarhus, Denmark}
\author{F. Oehl}
\author{R. C. Thompson}\email{r.thompson@imperial.ac.uk}
\affiliation{Quantum Optics and Laser Science, Blackett Laboratory, \\
 Imperial College London, Prince Consort Road, London, SW7 2AZ, United Kingdom
}


\begin{abstract}
We present the results of simulations of optical sideband cooling of atomic ions in a trap with a shallow potential well. In such traps, an ion cannot be Doppler cooled near to the Lamb-Dicke regime ($\eta^2(2\braket{n}+1)  \ll 1$). Outside the Lamb-Dicke regime, the sideband cooling dynamics are altered by the existence of various Fock states with weak coupling where the cooling becomes very slow. A $^{40}$Ca$^+$ ion trapped in our Penning trap realizes such a situation, hence single stage cooling is inefficient to prepare the ion in the motional ground state. For these systems,  it is necessary to study the cooling dynamics in detail and we show that it is possible to implement an optimized cooling sequence to achieve efficient ground state cooling. We also present the simulated cooling dynamics of two ions trapped in a Penning trap, where the presence of an additional motional mode requires a complicated cooling sequence in order to cool both axial modes to the ground state simultaneously. Additionally, we demonstrate the dissipative preparation of Fock states outside the Lamb-Dicke regime by sideband heating a single ion in a Penning trap.

\end{abstract}

\pacs{03.67.Lx, 37.10.Ty, 37.10.Rs, 42.50.Ct, 42.62.Fi}

\maketitle

\section{Introduction} Trapped ion systems are an ideal platform for realizing high fidelity quantum operations, due to their near-perfect isolation from the environment.  These systems are often used in studies such as quantum information processing, quantum simulations, precision measurements and studies of non-equilibrium dynamics \cite{Wineland_2003, Schmidt_2003, Monz_2016, Debnath_2016, Porras_2004, blatt_2012, An_2014}. Furthermore, laser cooling of trapped ions eliminates many uncertainties in the measurements. Good control over the motion of trapped ions, together with a near unity state preparation fidelity, makes these systems useful for taking repetitive measurements, hence reducing the statistical uncertainties \cite{Diddams_2001, bruun2011a}. Specifically, a Penning trap is very well suited for the confinement of large ion Coulomb crystals \cite{Mavadia_2013} due to the lack of micromotion, thus providing a large scale quantum simulator \cite{ Britton_2012}, and an open geometry allowing fundamental studies based on beamlines \cite{Heisse_2017, Ahmadi_2017}  to be carried out. Additionally, due to the absence of an RF field and the relatively large dimension of the trap electrodes, the anomalous heating rates in such traps are found to be low \cite{Goodwin_2016}. As a result, trapped ions in a Penning trap can be used to study interactions even when the required interrogation time is long.   

Controlled manipulation and precise measurement of a quantum state are prime requirements to study a large variety of interactions with high fidelity. An ion cooled to the motional ground state provides an ideal system in which high fidelity operations can be carried out. A narrow linewidth laser beam or microwave radiation can be used to manipulate the motional and electronic states of a trapped atomic ion \cite{Monroe_1995, Ospelkaus_2011}. Often, optical sideband cooling of trapped ions is realized by tuning a laser to a well-resolved red sideband of a narrow linewidth transition \cite{Diedrich_1989}, but ground state cooling is difficult to achieve in shallow traps. Such traps are characterized by a large value of the Lamb-Dicke (L-D) parameter \cite{Wan_2015, Goodwin_2016}, and a significant fraction of the population is found to be outside the L-D regime after Doppler cooling. The L-D parameter is the ratio of momentum provided by the photon and momentum required to change the motional state of the ion by unity. It thus determines the strength of motional sidebands in a resolved sideband spectrum of a trapped ion, and a large value results in several various resonance peaks in the spectrum. However a small value gives rise to only first order sidebands. Ground state cooling outside the L-D regime is challenging for three reasons: first, during the spontaneous emission of a photon, the motional state of the ion may change; second, during the cooling process, the population can be trapped in so-called minima where the coupling strength of the sideband is very weak; and third, off-resonant excitation to the carrier or the blue sidebands slows down the cooling (see Section IV). Some theoretical studies have previously been carried out to tackle the issues of sideband cooling outside the L-D regime \cite{Morigi_1997, Morigi_1999}. However, a detailed study of the cooling dynamics of a multi-level and multi-ion system has not yet been performed.

In this paper, we will present simulations of the sideband cooling of ions towards the motional ground state from far outside the L-D regime, and sideband heating from the ground state to high-lying motional states. Selected results will be reinforced by experimental data obtained with $^{40}$Ca$^+$ ions in our Penning trap. For this reason, we restrict our simulation studies to parameters readily achievable with $^{40}$Ca$^+$ in our setup, but the methods and qualitative conclusions apply to any other ion species with analogous level structure and to other traps operating outside the L-D regime.  The studies are important for systems where sympathetic cooling of different species is carried out, which otherwise cannot be laser cooled.  Here, in order to achieve a stable configuration, the system will initially lie outside the L-D regime. Furthermore, such systems can be used to amplify the recoil signal generated in recoil spectroscopy techniques \cite{Wan2014}.

\section{Motion of trapped ions and laser cooling} 
\subsection{Motion of a $^{40}$Ca$^+$ ion in a Penning trap} 
In this paper, we will study the axial motion (along the $z$ axis) of ions in a Penning trap. Confinement of ions in the axial direction is carried out by a static quadrupolar potential V = $U_{dc}(2 z^2-x^2-y^2)/{2d_0^2}$, where $U_{dc}$ is the applied potential and $d_0$ is related to the size of the trap. The motion in the axial direction is therefore a simple harmonic motion.    

In the radial plane ($xy$ plane) the ions are trapped with the help of a static axial magnetic field $\textbf{B} = (0,0,B_z)$ \cite{Major_2006, Mavadia_2013}. The motion in the radial plane is characterized by the modified cyclotron and magnetron orbital motions \cite{Ghosh_1995}. 
        
The characteristic angular frequencies of the axial mode ($\omega_z$) and radial modes ($\omega_+$, $\omega_-$) of a single trapped ion in a Penning trap are \cite{Ghosh_1995}
\begin{equation}
\omega_z =\sqrt{\frac{2 q U_{dc}}{m d_0^2}}, \hspace{0.2cm}  \omega_{\pm} =\tfrac{1}{2} (\omega_c \pm \sqrt{\omega_c^2 -2 \omega_z^2}),
\label{freq}
\end{equation}
where $m$ is the mass and $q$ is the charge of the ion. In the above equation, $\omega_c$ =$|q|B_z/m$ is the cyclotron frequency. The stability criterion for the motion in the radial plane, $\omega_z < \omega_c/\sqrt{2}$, limits the axial frequency for a given value of magnetic field strength \cite{Ghosh_1995}. Typically, for a singly charged $^{40}$Ca$^+$ ion in our trap (with a magnetic field of $\sim 1.9$ T) the axial frequency can take values up to 500 kHz for a corresponding 730 kHz cyclotron frequency. More details about the experimental apparatus can be found in \cite{Stutter_2017, Goodwin_2016}.

\subsection{Axial motion of a Coulomb crystal} 
At sufficiently low temperatures, ions in a trap will form a Coulomb crystal \cite{thompson2015}.  Out of 3$N$ orthogonal motional modes for an $N$-ion Coulomb crystal, at least $N$ can be interrogated with an axial laser beam. Our interest is to perform sideband cooling on the axial modes, and in this paper we restrict our studies to one and two ions.  

In a two-ion Coulomb crystal, the ions can take two different orientations with respect to the magnetic field axis. The first is a chain configuration where ions align themselves along the magnetic field direction and the second one is a planar configuration where the ions rotate with a constant separation in a plane perpendicular to the magnetic field. Depending upon the trapping force in the axial and radial directions, the ions can take one or the other orientation. In practice, the configuration can be changed by varying the trapping force and the parameters of the Doppler cooling lasers \cite{Mavadia_2013} (see next section). 

In any orientation of a two-ion Coulomb crystal, there are two orthogonal modes along the trapping axis.  The in-phase (center of mass) and out-of-phase (breathing) axial motional frequencies of a two-ion chain are given by $\omega_z$ and $\sqrt{3}\omega_z$, where $\omega_z$ is the motional frequency of a single ion as defined in equation \ref{freq}. The center of mass motion of a two-ion planar crystal is also at $\omega_z$. On the other hand, the out-of-phase (tilt) motion of a planar crystal depends upon the rotation frequency, which is set by the laser parameters as well as the trapping electric and magnetic field strengths. The motional frequency of the tilt mode is given by 
\begin{equation}
\omega_\text{tilt}=\sqrt{\tfrac{3}{2}\omega_z^2-\omega_r (\omega_c - \omega_r)},
\label{freqT}
\end{equation}
where $\omega_r$ is the rotation frequency of the crystal in the radial plane.

\subsection{Laser cooling} 
 Doppler cooling is now a standard technique to cool trapped atomic ions and optimum cooling is achieved with an unsaturated laser beam that is red detuned by half of the transition linewidth from the required transition \cite{Wineland_1978}. In a Penning trap,  laser cooling in the radial direction needs additional features to overcome the effect of a deconfining electric force. Laser cooling of the modified cyclotron motion is carried out with a red-detuned beam in the radial plane. Additionally, cooling of the magnetron motion is realized by offsetting the beam from the trap center in order to produce a gradient of laser intensity across the trap center \cite{Itano_1982}. In some cases an additional time-varying axialization field at $\omega_c$ \cite{Phillips_2008}, which couples the magnetron motion to the modified cyclotron motion, leads to stronger Doppler cooling of radial modes for very small numbers of ions. 

\begin{figure}[ht]
\centering
\includegraphics[width=0.5\textwidth, trim={1cm 2.5cm 1cm 1cm},clip]{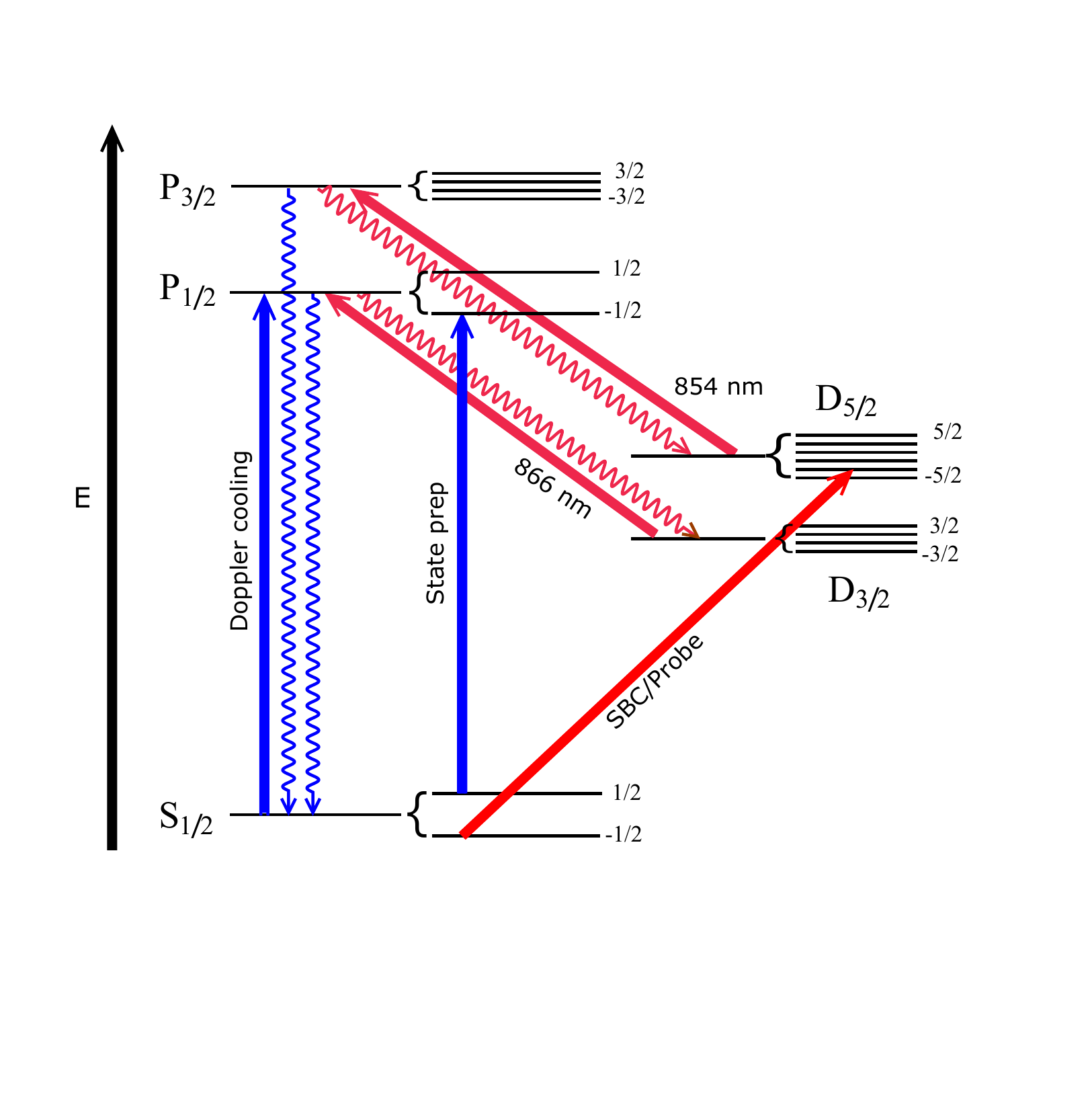}
\caption{A partial energy level diagram for a $^{40}$Ca$^+$ ion in the presence of a magnetic field. Doppler cooling is carried out on the S$_{1/2} \leftrightarrow \text{P}_{1/2}$ transition (at 397 nm). Spontaneous decay to the D$_{3/2}$ and magnetically induced decay to the D$_{5/2}$ levels interrupt the laser cooling. These levels are optically repumped to continue the laser cooling with the help of lasers at $866$ nm and $854$ nm. The narrow transition  S$_{1/2} \leftrightarrow \text{D}_{5/2}$ (at 729 nm) is used for sideband cooling.}
\label{CaLevel}
\end{figure}

Doppler cooling is generally carried out between atomic levels where the linewidth of the transition is much greater than the trap frequency, so the motional sidebands are unresolved. In the $^{40}$Ca$^+$ ion system, the Doppler cooling is carried out on the $\text{S}_{1/2} \leftrightarrow \text{P}_{1/2}$ transition. Due to the presence of a large magnetic field in a Penning trap, the electronic levels are subject to a correspondingly large Zeeman splitting.  Hence,  multiple lasers are needed to implement laser cooling. A partial level scheme of a $^{40}$Ca$^+$ ion in the presence of a magnetic field is shown in figure \ref{CaLevel}. The final temperature achieved after Doppler cooling in the axial direction is roughly 0.5 mK \cite{Goodwin_2016}. In our simulations, we assume an initial mean phonon number $\langle n \rangle$ corresponding to the Doppler limit expressed in terms of a thermal phonon distribution function \cite{Leibfried_2003}.

\subsection{Resolved sideband cooling} 
Resolved sideband cooling is performed on the S$_{1/2}\leftrightarrow$ D$_{5/2}$ transition of a $^{40}$Ca$^+$ ion at 729 nm. In our simulation, we include the effect on the sideband cooling arising from the presence of multiple levels. The lifetime of the D$_{5/2}$ level is of the order of 1 s, after which the ion spontaneously decays to the S$_{1/2}$ level \cite{Barton_2000}. In order to achieve an efficient sideband cooling rate, the population from the D$_{5/2}$ level is repumped to the P$_{3/2}$ level through an off-resonant 854 nm laser, from where it predominantly decays back to the S$_{1/2}$ level. The repumping rate can be set by the 854-nm laser parameters, such as intensity and detuning from the transition. For efficient cooling, the repumping rate needs to be optimized. We discuss the required 854-nm repumping and 729-nm excitation rates in the next section. 
    
The population in the P$_{3/2}$ state decays to both Zeeman sublevels of the S$_{1/2}$ state and a part of the population needs to be excited to the P$_{1/2}$ state in order to return it to the correct Zeeman sublevel for sideband cooling (i.e. $m_j=-1/2$ in our case). This leads to an additional spread of the population in the motional Fock states because of the large L-D parameter for this transition. Due to the multiple levels involved, the cooling slows down and also the final ground state occupancy is reduced. In order to simulate the system, ideally, we would need all the transition rates and laser parameters, however we can neglect the time required for the S$_{1/2}$ to P$_{1/2}$ transition as it takes place on a timescale of a few  ns and the main effect will be additional heating during this transition. To include the heating effects, we find a normalized diffusion coefficient for all the respective branches which takes into account the amount of momentum transfer during each transition. Table 1 shows the parameters for which the simulation is carried out. 

\begin{table}[ht]
\caption{Typical transition rates used for the simulation \cite{Barton_2000}.}

\begin{tabular}{ |c||c|} 
\hline
$\Gamma_{854}$ =   $2\pi \cdot 1.6$ MHz  & $\tilde{\Gamma}$ =  $2\pi \cdot 40$ kHz \footnote{This is the effective repumping rate set by the intensity and detuning of the 854 nm laser.}\\
\hline
$\Gamma_{866}$ =  $2\pi \cdot1.3$ MHz  &$\Omega_{729}$ = $ 2\pi \cdot 40$ kHz  \\ 
\hline 
$\Gamma_{393}$ =  $2\pi \cdot 21.5$ MHz  & $\Gamma_{729}$ = $ 2 \pi \cdot 0.15$ Hz  \\ 
\hline
$\Gamma_{397}$ =  $2\pi \cdot 21.2$ MHz  & $\omega_z$ =  $2\pi \cdot$ 100 - 500 kHz  \\ 
\hline
\end{tabular}
\end{table}

\section{Theory of sideband cooling of trapped ions}
\subsection{Interaction of laser light with a two-level atom} 
The Hamiltonian for a two-level atom in the presence of a classical light field can be written as 
\begin{equation}
H =\frac{\hbar \omega_0}{2} \sigma_z+  \frac{\hbar \Omega_0}{2} (e^{i( \omega_l t-k z)}+e^{-i(\omega_l t-k z)})(\sigma_+ + \sigma_-), 
\end{equation}
where the first term is the free atom Hamiltonian and the second term is  the interaction part of the Hamiltonian. $\omega_0$ is the angular frequency of the transition between the electronic ground state $\ket{g}$ and excited state $\ket{e}$,  $\Omega_0$ is the Rabi frequency and $\omega_l$ is the frequency of the monochromatic radiation. The operators $\sigma_+$ and $\sigma_-$ are the raising and lowering operators for the internal states of the atom, respectively \cite{Mandel_1995}.

\subsection{Interaction Hamiltonian in the presence of a trapping force}
The Hamiltonian corresponding to the axial motion of a single ion in the presence of harmonic trapping fields can be written as
\begin{eqnarray}
H_\text{trap} = \hbar \omega_z(a^{\dagger} a +1/2),
\end{eqnarray}  
where  $a^{\dagger}$ and $a$ are the raising and lowering operators for the motion. The expression can be further simplified by transforming to the interaction picture. After making the rotating wave approximation, the Hamiltonian can be written as \cite{Leibfried_2003}  
\begin{equation}
{{H}}_\text{int} = \frac{\hbar \Omega_0}{2} \left(  \sigma_+   e^{-i\eta(\tilde{a}^{\dagger} + \tilde{a} )} e^{-i\Delta t} \right) + \text{h.c.},
\end{equation}
where  $\Delta$ is the laser detuning from the transition. The new raising and lowering operators in the interaction picture are defined as $\tilde{a}^{\dagger} = a^{\dagger}e^{-i\omega_z t}$ and $\tilde{a} = a\ e^{i \omega_z t}$. The parameter $\eta$ is called the L-D parameter and is related  to the trapped ion oscillation frequency by $\eta =\sqrt{\hbar k^2/(2 m \omega_z)}$. This parameter determines how strongly the motional sidebands can be excited during optical spectroscopy.  

The energy levels of the trapped ion split into multiple sub-levels spaced by the motional frequency $\omega_z$. The coupling strength between the internal states $\ket{g}$, $\ket{e}$ and motional states $\ket{n}$, $\ket{n'}$ can be expressed in terms of the Rabi strength 
\begin{eqnarray} \label{omega_nn}
\Omega_{n,n'} &=& \Omega_0  { \bra{n'} e^{i\eta (\tilde{a}+\tilde{a}^\dagger)}\ket{n}}, \\ \nonumber
&=& \Omega_0  \sqrt{\frac{n_<!}{n_>!}}\eta ^{\left| \Delta n\right| } e^{-\frac{\eta ^2}{2}} L_{n_<}^{\left| \Delta n \right| }\left(\eta ^2\right),
\label{Rabi_Strength} 
\end{eqnarray} 
where $L_{n}^{\left| \Delta n \right| }(x)$ is the generalized Laguerre polynomial and $\Delta n = n' -n$. The parameters $n_>$ and $n_<$ correspond to the higher and lower values of the phonon states involved in the optical excitation from state $\ket{g,n}$ to $\ket{e,n'}$.
 
\subsection{Interaction of laser radiation with two ions} 
The L-D parameters corresponding to the normal modes of a two-ion crystal differ from the single-ion one due to the difference in the collective mass of the crystal and the collective motional frequencies, and are given by, $\eta_\text{COM} =\eta/\sqrt{2}$  and $\eta' = \eta/\sqrt{2 \omega'/\omega_z}$ corresponding to the COM and breathing (or tilt) motional modes, respectively. Here $\omega_z$ and $\omega'$ are the COM and breathing (or tilt) motional mode frequencies. The interaction Hamiltonian of a two-ion crystal is expressed as
\begin{equation}
\begin{split}
{{H}}_{I} = \frac{\hbar  \Omega_0}{2}  \sum_{n,n'} \sum_{m,m'} \sum_{j=1}^{2}\sigma_+^{(j)}  e^{-i\Delta t}  \big[ \ket{n'}\bra{n}\tilde{\Omega}_{n',n}e^{-i(n'-n)\omega_z t} \\
\times \ket{m'}\bra{m}\tilde{\Omega}_{m,m'}e^{-i(m'-m)\omega' t} (-1)^j \big] + \text{h.c.},
\end{split}
\end{equation}
where the ions are labelled by $j=1,2$.  The Rabi frequency of the motional excitation for both of the motional modes ($n\rightarrow n'$ for COM, $m \rightarrow m'$ for breathing/tilt) is written as  
\begin{equation} 
\Omega_{n,n',m,m'}= \Omega_0 \cdot \tilde{\Omega}_{n,n'} \cdot \tilde{\Omega}_{m,m'}
\end{equation}
where $\tilde{\Omega}_{n,n'} =\Omega_{n,n'}/\Omega_{0}$ is the normalized Rabi strength of the transition from motional state $n$ to $n'$ (see equation \ref{omega_nn}), and similarly for $\tilde{\Omega}_{m,m'}$. 

\subsection{Rate equation approach for resolved sideband cooling}
 The population dynamics of an atomic system can be described in terms of a master equation, which is given by
\begin{equation}
\frac{d\rho(t)}{dt}  = -\frac{i}{\hbar} \left[ V(t), \rho(t) \right]+\mathcal{L}\rho(t),
\end{equation}
where $\rho(t)$ is the population density matrix at time $t$ and $V(t)$ is a time dependent interaction term, set by the laser intensity and detuning. Often, the cooling involves multiple levels. For simplicity the spontaneous emission between these levels can be defined in terms of the Liouvillian operator $\mathcal{L}$ and is expressed as  
\begin{equation}
\mathcal{L}\rho = \frac{\Gamma}{2} (2 \sigma_{-} \tilde{\rho} \sigma_{+} -  \sigma_{+}\sigma_{-} \rho - \rho \sigma_{+} \sigma_{-})
\end{equation}
where $\Gamma$ is the transition rate from the excited state $\ket{e}$ to a third level (say from D$_{5/2}$ to P$_{3/2}$ in the case of a $^{40}$Ca$^+$ ion) and $\tilde{\rho}$ is a modified density matrix that accounts for the change in momentum during spontaneous emission \cite{Stenholm1986}.  

 By reducing the multi-level atom to a two-level equivalent and  taking the motional states into account, the population dynamics between external states of a single ion can be written as \cite{marzoli_1994} 
\begin{align}
\frac{d P_{n}}{dt}  &=  -\sum_{k}\frac{\tilde{\Gamma} \text{$\Omega^2_{n,k} $}}{\tilde{\Gamma}^2+4 (\delta +\text{$ (n-k) $} \text{$\omega _z$})^2+ 2\text{$\Omega^2_{n,k} $}}  P_n \nonumber \\ 
& + \sum_{r }\sum_{k}\frac{\tilde{\Gamma}\text{$\Omega^2_{r,k} $}}{\tilde{\Gamma} ^2+4 (\delta +\text{$ (r-k) $} \text{$\omega _z$})^2 +2\text{$\Omega^2_{r,k} $}} D_{k \rightarrow n}P_r
\label{Pn}
\end{align}
where $\tilde{\Gamma}$ is the effective repumping rate, which can be written in terms of the repumping laser Rabi frequency ($\Omega_{854}$) and detuning from the transition ($\delta_{854}$) 
\begin{equation}
\tilde{\Gamma} =\frac{ \text{$\Omega_{854}^2 $}}{ 4\delta_{854}^2+(\text{$\Gamma_{393} $}+\text{$\Gamma_{854} $})^2} \text{$\Gamma_{393}$}.
\label{Effective_Rate}
\end{equation}
The term $D_{k \rightarrow n}$ is called the diffusion function which accounts for the population spread between the motional states during the spontaneous emission of photons and is expressed as 
\begin{eqnarray}\label{eq:DiffAll}
\mathcal{D}_{k \rightarrow n} = \sum_{p} \sum_{m} \sum_{l} |\tilde{\Omega}_{k,l}(\eta_{854a})|^2 |\tilde{\Omega}_{l,m}(\eta_{393d})|^2\\ \nonumber \times |\tilde{\Omega}_{m,p}(\eta_{397a})|^2 |\tilde{\Omega}_{p,n}(\eta_{397d})|^2, 
\end{eqnarray}
 where the L-D parameters account for a four step diffusion process consisting of absorption at 854 nm ($\eta_{854a}$), decay at 393 nm ($\eta_{393d}$) into the incorrect $S_{1/2}$ sublevel followed by absorption ($\eta_{397a}$) and decay ($\eta_{397_d}$) at 397 nm that completes the cooling cycle.  This simplified model stems from the fact that each transition occurs on average once when accounting for the different transition probabilities given by the Clebsch-Gordan coefficients. The summations over $p, m$ and $l$ take into account all motional states to which the diffusion could take place during sideband cooling.

The equivalent expression has also been used in reference \cite{Lindberg1984}. In order to resemble our experimental conditions in \cite{Stutter_2017, Goodwin_2016} the diffusion function is normalized to include the equal absorption rate due to the radial and axial blue laser beams (397 nm) and the absorption of the repumping laser (854 nm) is assumed to be in the axial direction only. 

The 854 nm laser shifts the D$_{5/2}$ level up or down in energy depending on the sign of the detuning from the transition. An expression for the Stark shift ($\Delta_\text{AC}$)  can  be formulated for the $^{40}$Ca$^+$ ion system \cite{marzoli_1994} and is written in terms of the 854 nm detuning ($\delta_{854}$) and the Rabi frequency ($\Omega_{854}$) of the transition 
\begin{equation}
\Delta = \frac{\delta_{854} \Omega_{854}^2 }{ 4 \delta_{854}^2+(\Gamma_{393} +\Gamma_{854} )^2}
\label{Stark_Shift}
\end{equation}
where $\Gamma_{393}$ and $\Gamma_{854}$ are the linewidths of the S$_{1/2}\leftrightarrow $ P$_{3/2}$ and D$_{5/2}\leftrightarrow$ P$_{3/2}$ transitions respectively. Typically for our system the Stark shift is set to a few tens of kHz and the repumping rate is typically $30$ to $40$ kHz. 

Unlike previous studies in references \cite{marzoli_1994, Vogt1996}, the rate equation approach described here (as given in equation \ref{Pn}) is not restricted to the L-D regime. The analytical solution of the rate equation is not possible due to the complicated diffusion function; hence a numerical simulation is carried out.

\subsection{Monte Carlo approach to simulate cooling dynamics}
A Monte Carlo approach is often a suitable alternative when the numerical solution of the rate equations is computationally too expensive, particularly in multiple dimensions. For this purpose we need to model the multi-level system and find the probabilities corresponding to each expected transition. A random number generator can be assigned to simulate the cooling dynamics. As the transitions in atomic systems take place through discrete jumps, solving such systems via a Monte Carlo approach resembles  the actual experiments quite well. 

In this paper we will discuss the results obtained using both methods. For the single ion case, we mainly numerically integrate the rate equations directly but in the multi-ion case this approach is inefficient, hence the Monte Carlo method is adopted. 

Before the start of the simulation, the transition rates for all the applicable decay and excitation branches of all the included Fock states are precomputed and stored as a matrix corresponding to each state. The simulation then proceeds as follows:

1. The system is initialized into a phonon state sampled from the Doppler-cooled Fock state distribution. 

2. Uniform random numbers are sampled to check for the sideband cooling excitation followed by the composite repumping  transition. 

3. The diffusion coefficients are sampled to find the phonon number change during repumping and spontaneous decay processes. 

4. The time is updated for each transition. 

5. Steps 2--4 are repeated the desired number of times corresponding to the cooling length being simulated to give the final Fock state. 

6. Steps 1--5 are repeated 10000 times.

\section{Results and discussion}
Theoretical studies of resolved sideband cooling within and outside the L-D regime have been carried out previously \cite{Morigi_1997, Morigi_1999}. Especially outside the L-D regime, the studies were limited to a two-level atomic system starting with a small mean phonon number. In this paper we discuss the resolved sideband cooling of $^{40}$Ca$^+$ ions where the population starts far outside the L-D regime. The simulation also considers the impact on sideband cooling due to the additional atomic sub-levels arising from the large magnetic field required for a Penning trap, which slows down the cooling due to the diffusion of the population into different phonon number states during the transitions between these additional sub-levels.  

We therefore extend the simulation studies from a two-level system to a five-level system in a Penning trap. The effect of additional levels on the sideband cooling is separately included in the form of a diffusion coefficient. Our approach assumes that the rate-limiting steps are the excitation rates on the 729 and 854 nm transitions and all other decay and repumping transitions are orders of magnitude faster and thus can be treated as instantaneous. We also extend our studies to multiple ions in a small Coulomb crystal, which has not been investigated thoroughly in the previous studies. In our simulation, we do not consider any effect due to the coherence or entanglement between atomic and motional states because these effects do not play any significant role for the cooling parameters stated above. 

\subsection{Sideband cooling of a single ion} 
We initially discuss the case of a single trapped $^{40}$Ca$^+$ ion in a Penning trap. In the simulation, we assume that the ion is cooled to its Doppler limit. The mean phonon number at the Doppler limit for a range of frequency from $100$ kHz to $500$ kHz is found to be between 95 and 19. The L-D parameter on the S$_{1/2} \leftrightarrow$ D$_{5/2}$ transition (729 nm radiation) lies between 0.31 and 0.14. Note that for the above frequency range, the system lies far outside the L-D regime after Doppler cooling. 

\subsubsection{Optimization of cooling parameters} 
We optimize the parameters by simulating the cooling dynamics for a range of sideband cooling laser and repumping laser parameters. We consider all the involved atomic levels as discussed in the previous section. The sideband cooling and repumping lasers are assumed to be aligned along the trap axis. Diffusion due to the decay at 393 nm is assumed to be in all directions, and a corresponding L-D parameter from the associated momentum transfer in the direction of motion is calculated.

In order to complete the cooling cycle, it is necessary to prepare the ion in the correct ground state sublevel S$_{1/2,-1/2}$, hence an additional laser at 397 nm is necessary to recycle the ion if it falls into the other sublevel S$_{1/2,1/2}$. We also include the diffusion of the population due to re-emission at 397 nm, again in all directions. In the end, we calculate the mean phonon number corresponding to a given cooling time and laser parameters. 

\begin{figure}[h!]
\centering
\includegraphics[width=0.47\textwidth]{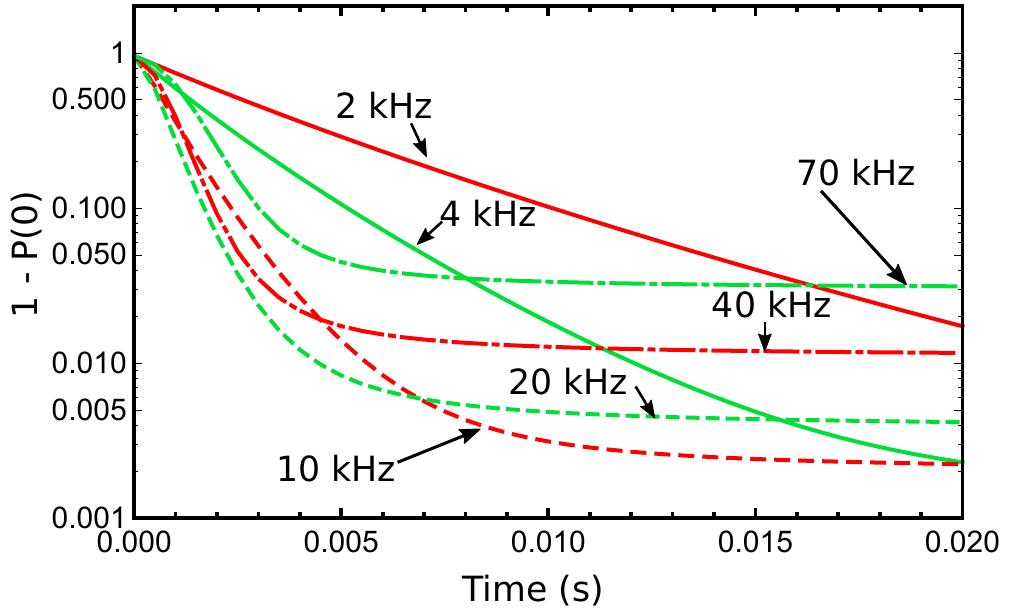}
\caption{Simulated population outside the $n = 0$ state as a function of cooling time for different repumping rates ($\tilde{\Gamma}/2 \pi$). The simulations are carried out at a motional frequency of $420$ kHz and Rabi frequency of $40$ kHz.}
\label{854_DetVsCool}
\end{figure}

At first we carry out the simulation at a motional frequency of $420$ kHz ($\eta_{729}$ = 0.15). At this frequency the effects of being outside the Lamb-Dicke regime are not large. The initial mean phonon number following  Doppler cooling is 22. The simulated population in the $n = 0$ state after sideband cooling, at different effective repumping rates, is shown in figure \ref{854_DetVsCool}. In practice the effective repumping rate can be varied by changing $\Omega_{854}$ or $\delta_{854}$ in equation \ref{Effective_Rate}. The Rabi frequency of a free atom is taken to be $2 \pi \times 40$ kHz and the detuning is set to the Stark shifted detuning, as given by equation \ref{Stark_Shift}.

The results show that the highest ground state occupancy can be achieved with a combination of longer cooling times and an effective repumping rate low enough to limit off-resonant excitation, but sufficient to overcome any heating processes. If fast cooling rates are required, then the repumping rate should be at least $20$ kHz, but one can only achieve an occupation probability of 99.5\% at this repumping rate. In order to achieve a lower mean phonon number one has to reduce the repumping rate. Realistically, a ground state occupation up to 99.7\% can be achieved after 20 ms of cooling at a repumping rate of $\sim 10$ kHz.


\subsubsection{Sideband cooling outside the L-D regime} 
\begin{figure}[ht]
\centering
\includegraphics[width=0.42\textwidth]{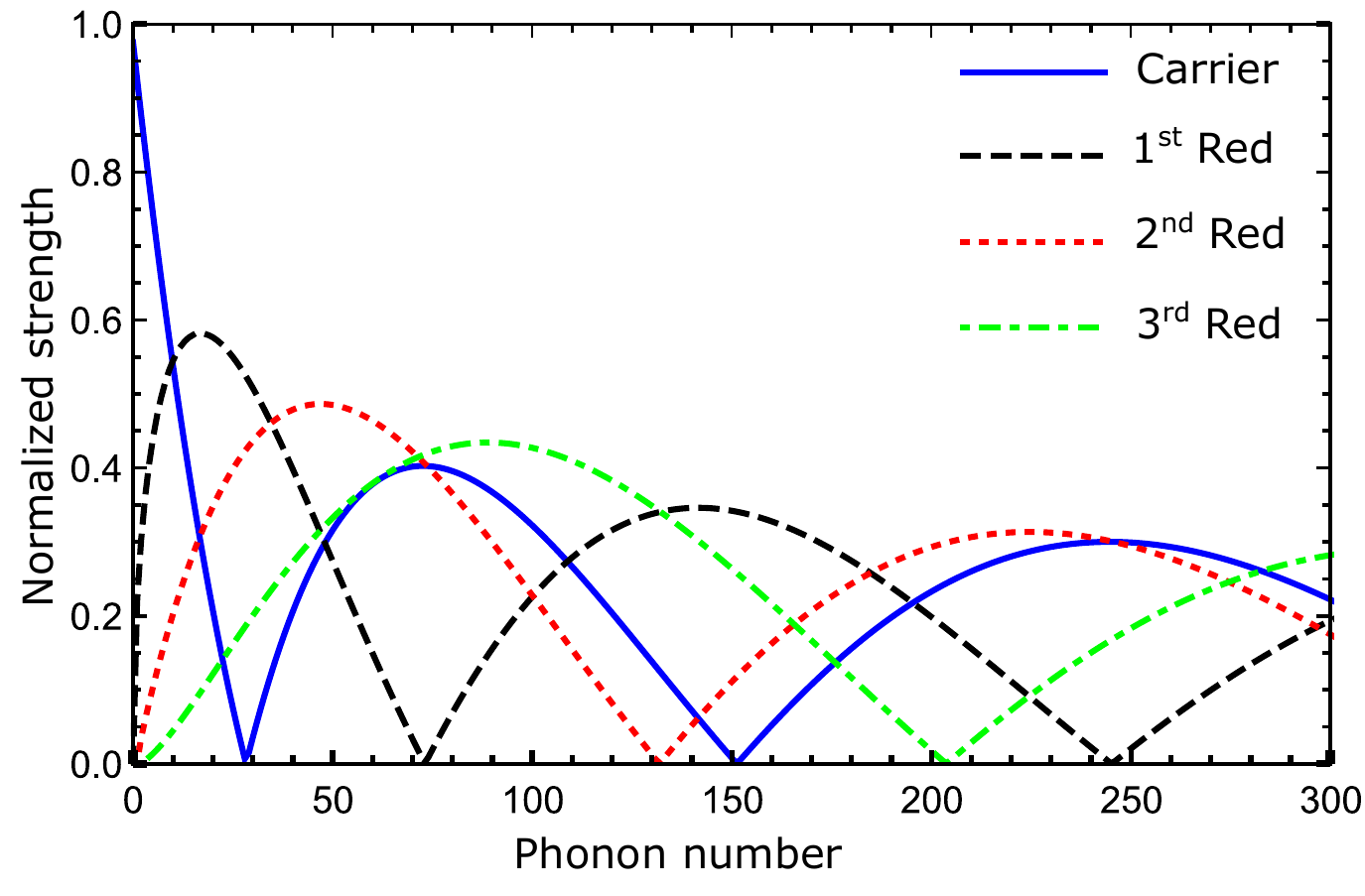}
\caption{Calculated sideband strengths for individual phonon number states at $187$ kHz ion oscillation frequency. The carrier is shown as a solid blue line, the first red sideband is shown in dash-black, the second red sideband in dot-red and third red sideband is shown in dot-dash green. }
\label{SBCStrength187k}
\end{figure}

Cooling of trapped ions outside the L-D regime is complicated by the fact that the strength of the sidebands varies for different phonon numbers. This can be noted from figure \ref{SBCStrength187k}. We choose the motional frequency to be $187$ kHz, the repumping rate to be $20$ kHz and the Rabi frequency of the transition is assumed to be $40$ kHz. The L-D parameter at this frequency is $\eta_{729} = $0.224 with a mean phonon number after Doppler cooling of 47. For these parameters the first red sideband strength approaches zero around $n\sim 73$. The cooling rate therefore is found to be close to zero for the population in this state and the states nearby. Eventually, population trapping in these phonon number states leads to a poor ground state occupation after sideband cooling \cite{Stenholm1986, Morigi_1997}. This population trapping can be avoided by cooling on higher order sidebands \cite{Goodwin_2016, Poulsen_2012}.

\begin{figure}[ht]
\centering
\includegraphics[width=0.23\textwidth]{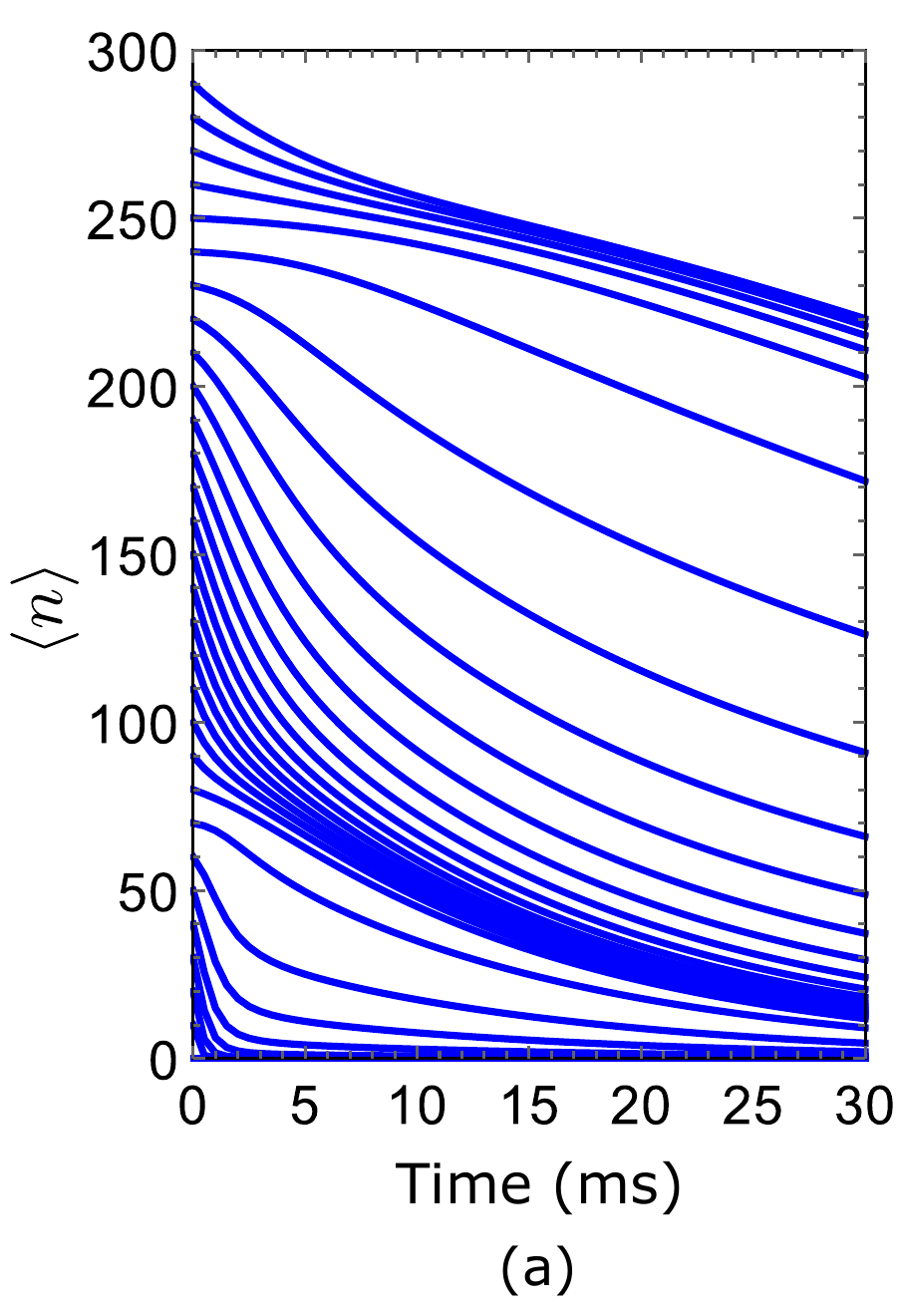}
\includegraphics[width=0.23\textwidth]{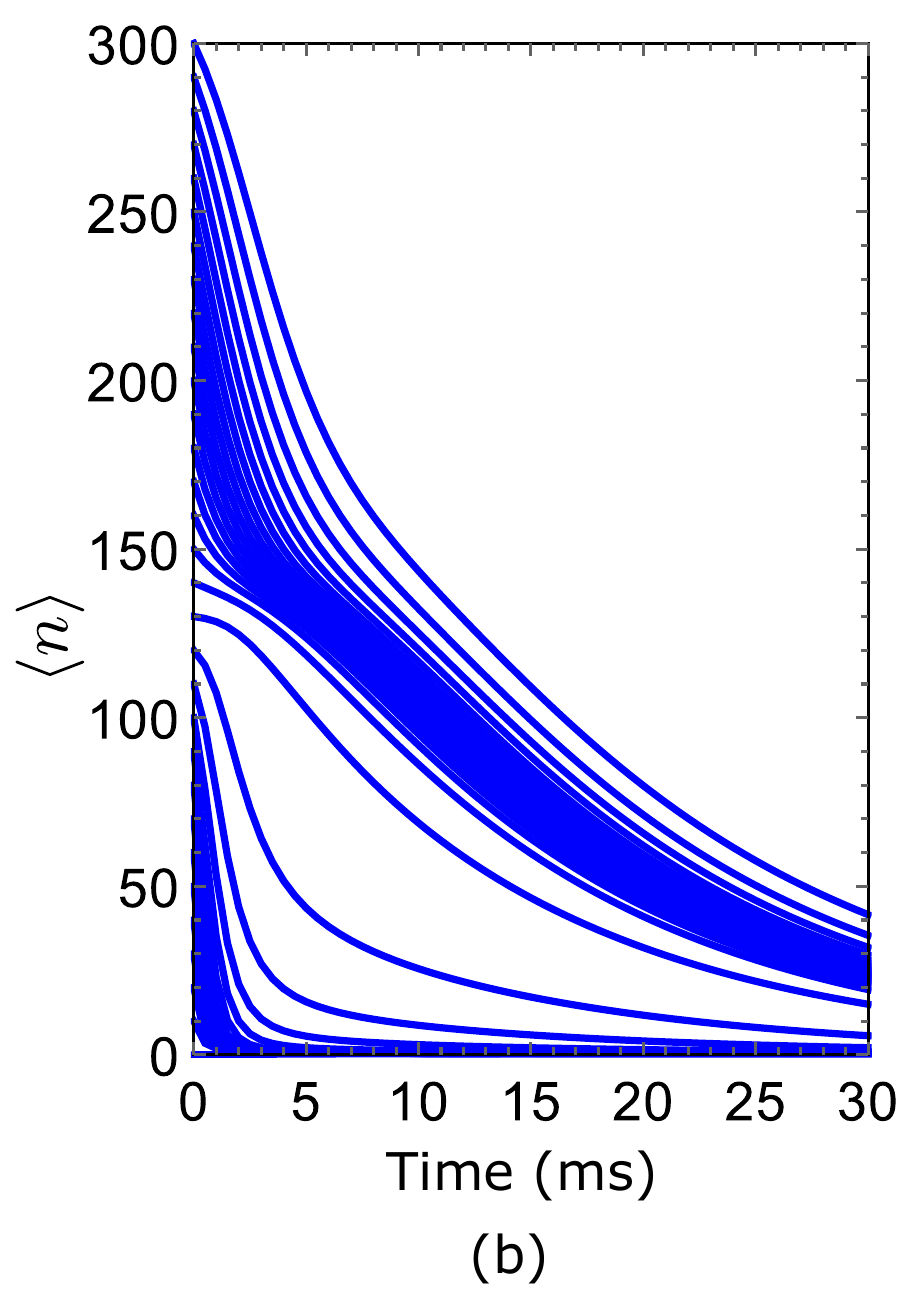}
\caption{(a) Simulated mean phonon number after single stage sideband cooling on the first red sideband for population initialized to a fixed Fock state number. (b) The same for the second red sideband.}
\label{CoolingDyna}
\end{figure}

Population dynamics of a single ion for aforementioned trap and laser cooling parameters have been simulated using a rate equation approach and the results are shown in figure \ref{CoolingDyna}. 
In the simulation, the population is initialized in a fixed Fock-state number and the mean phonon number is calculated for a given sideband cooling time. It can be seen from figure \ref{CoolingDyna} (a) that the population around $n\sim 73$ is cooled very slowly.  This is undesirable since time spent carrying out sideband cooling reduces the duty cycle of experiments. Furthermore, when heating rates are accounted for, this trapped population may never reach the ground state. In order to predict the behavior of higher order sidebands, we also simulated the cooling dynamics for the second order sideband (see figure \ref{CoolingDyna} (b)). We see that any population trapped at the first sideband minimum can be cleared out by implementing cooling on the second order sideband. This is possible due to the disparity in sideband minima position in both cases. More importantly, distinct sideband positions allow us to always choose a cooling sequence which deterministically  avoids any sort of population trapping.

As we have seen, population trapping is one of the features of systems outside the L-D regime. This population trapping gets more severe with decreasing motional frequency, i.e. increasing L-D parameter. The population trapped above the first zero of the first red sideband can be up to 65\% at $100$ kHz motional frequency, hence, in this regime we need a cooling sequence which can eliminate any sort of population trapping. 

In figure \ref{SBCStrength187k}, the strengths of the first, second and third red sidebands are shown for a motional frequency of $187$ kHz. It can be seen that the sideband strengths always have minima at different Fock states, hence switching between sidebands is effective to avoid any population trapping. Additionally, cooling on the higher order sidebands is beneficial due to the fact that each photon absorption event will remove more than one phonon from the system, thus cooling is fast. However, for low lying phonon states the higher order sideband strength drops drastically, thus making it inefficient. This motivates us to choose a cooling scheme where higher order sidebands are used at initial times when high lying Fock states are predominantly occupied and then switch to lower order ones as the population transfers towards lower Fock states. 


 
\begin{figure}[ht]
\centering
\includegraphics[width=0.42\textwidth]{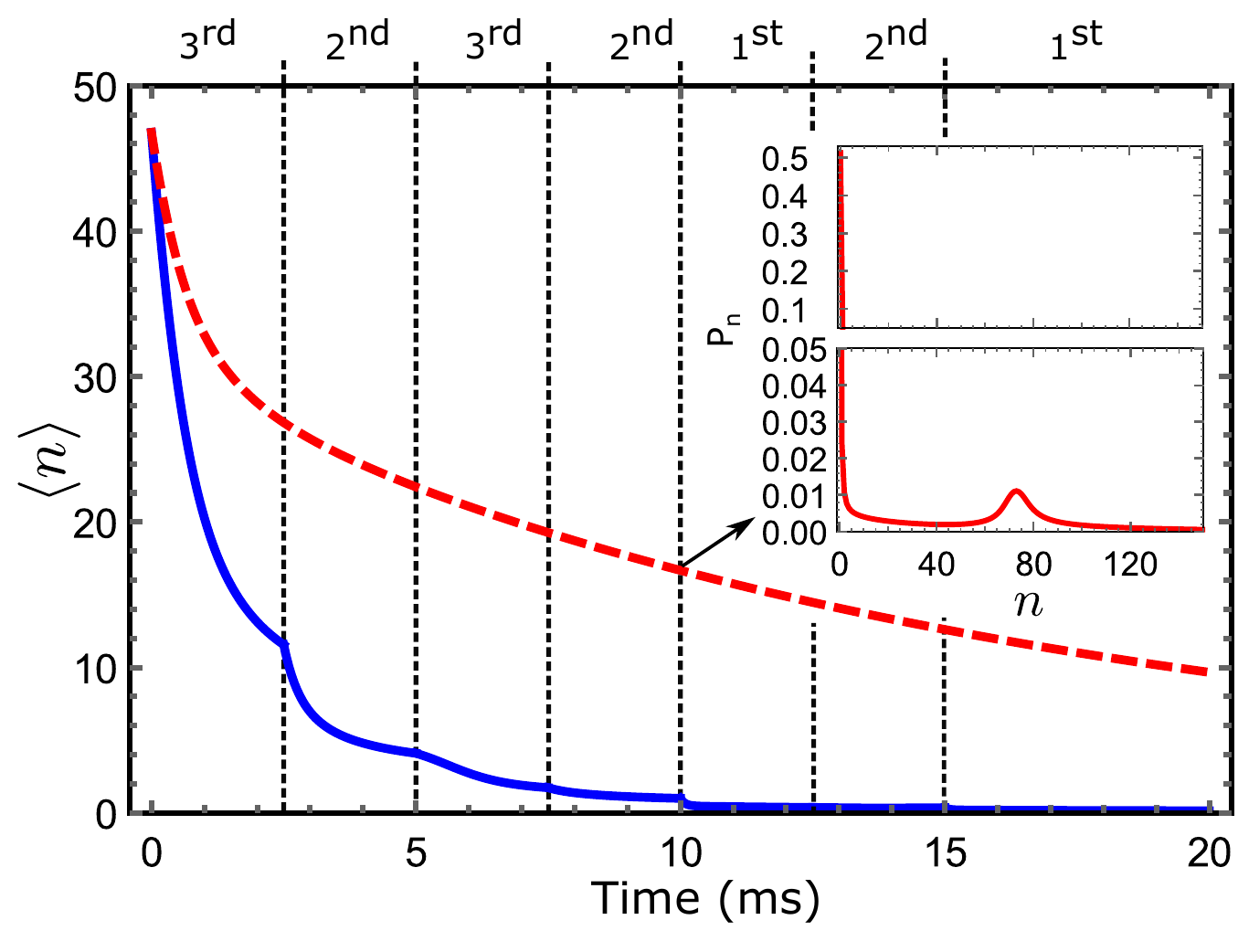}
\caption[Simulation of SBC at 187 kHz]{Simulated cooling dynamics of a single ion at 187 kHz motional frequency. The curve in red/dashed corresponds to single stage cooling (cooling with first red sideband) and the curve in blue/solid corresponds to multi-stage cooling. The pulse sequence is shown at the top of the frame. Population distribution after 10 ms of single stage cooling is shown in the inset.
}
\label{fig:187kSBCSim}
\end{figure}

 Based upon the qualitative discussion of the cooling dynamics results presented above, we formulate a cooling sequence that efficiently cools the ion to the ground state. However, this necessarily does not infer that the cooling is fastest. A simulated curve of the mean phonon number as a function of cooling time is shown in figure \ref{fig:187kSBCSim} for a Doppler cooled $^{40}$Ca$^+$ ion. The fraction of the population above the first sideband minimum point is around 22\%. The curve in blue/solid corresponds to multi-stage cooling and the curve in red/dashed corresponds to single-stage cooling. Clearly, the simulations show near ground state cooling for the multi-stage sideband cooling sequence. In this case, we choose a cooling sequence 3\textsuperscript{rd}-2\textsuperscript{nd}-3\textsuperscript{rd}-2\textsuperscript{nd}-1\textsuperscript{st}-2\textsuperscript{nd}-1\textsuperscript{st}, where cooling is performed for 2.5 ms at each sideband and 5 ms for the last pulse. The cooling sequence is chosen from the knowledge of the sideband strengths.

\begin{figure}[ht]
\centering
\includegraphics[scale=0.48]{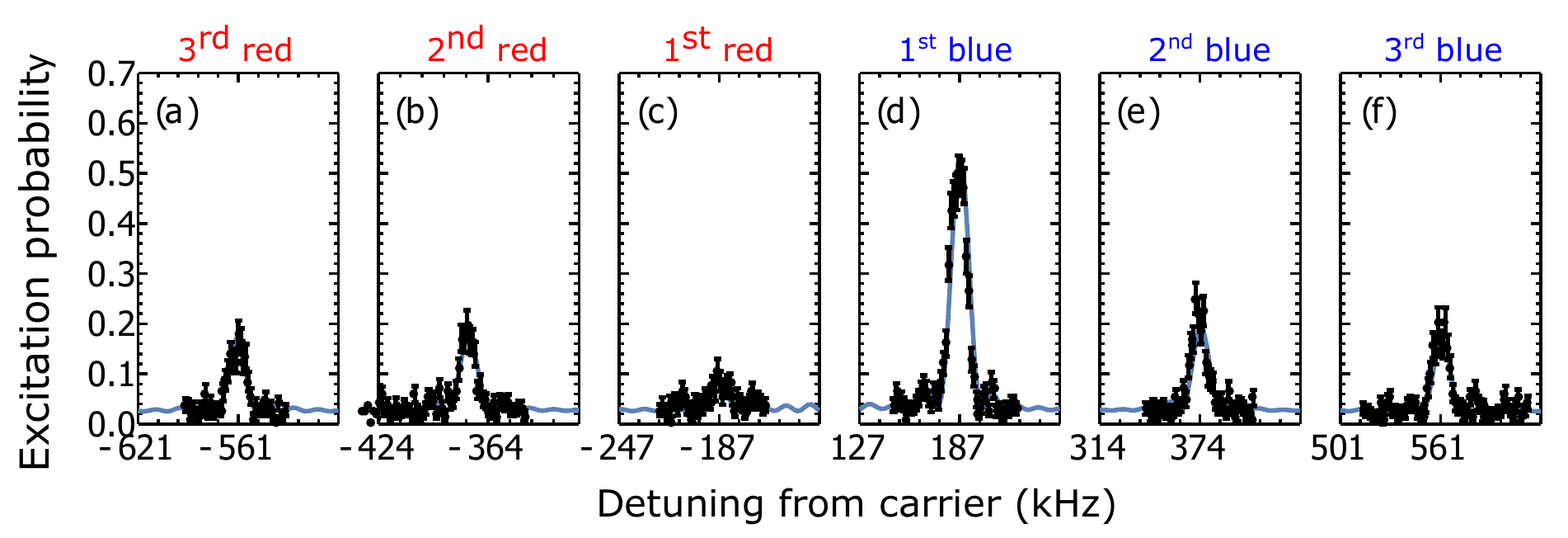}
\caption{Figure showing a spectrum after 10 ms of resolved sideband cooling on the first red sideband. The first red sideband (c) is nearly zero, however, the second and third red sidebands (a, b) show excitation indicating some population is trapped near the first red sideband of the transition. (d, e, f) show the 1$^\text{st}$, 2$^\text{nd}$ and 3$^\text{rd}$ blue sidebands respectively  }
\label{fig:SingleStageSBC}
\end{figure}

\begin{figure}[ht]
\centering
\includegraphics[scale=0.65]{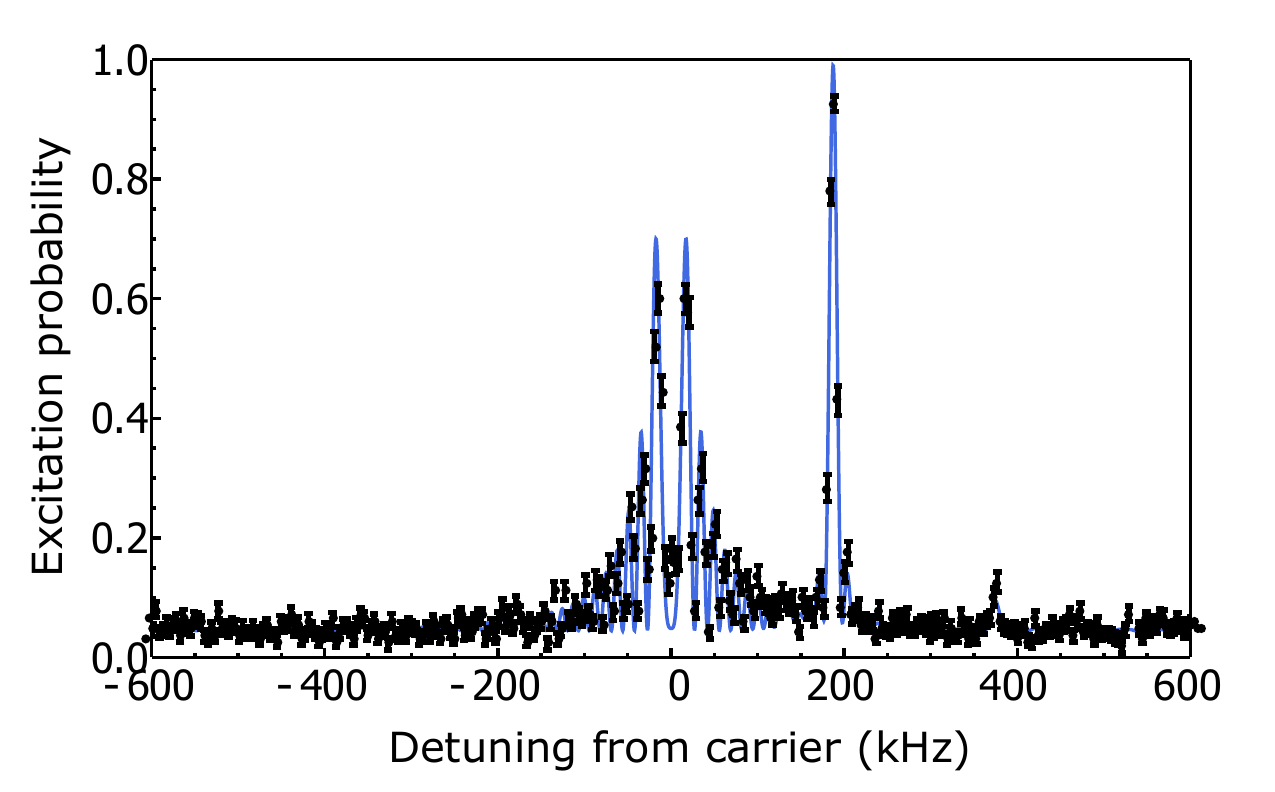}
\caption {A spectrum after  multi-stage sideband cooling of the axial motion at 187 kHz trap frequency. The mean phonon number is measured to be $\braket{n} = 0.03 \pm 0.01$ after 20 ms of sideband cooling.}
 \label{fig:MultiStageSBC}
\end{figure}

For qualitative comparison, experiments have been performed in our laboratory for the same trapping and laser conditions that are used in the simulations. The experimental apparatus and methodology have been discussed in reference \cite{Stutter_2017}. After Doppler cooling, the ion is sideband cooled for a total of 20 ms and then a sideband spectrum is acquired. The probe time (time for which the spectroscopy laser is turned on) is set to be a $\pi$ pulse for the first blue sideband. In figure \ref{fig:SingleStageSBC}, sideband cooling is performed on the first red sideband of the transition. The fitting estimates that $ \sim 84 \%$ of the population is very near the ground state with $\bar{n}=0.08(2)$ and the remaining $16 \%$ is trapped around the first red sideband minimum. 
The spectrum also shows that the excitation probability for the first red sideband is zero but at the same time, the second and third red sidebands are visible, which confirms that some of the population is trapped at the first red sideband minimum point. However, after implementing a multi-stage cooling sequence the population trapping is avoided and approximately 97\% of the population is cooled to the ground state with no appreciable population remaining near the trapped state. Corresponding experimental results are shown in figure \ref{fig:MultiStageSBC}.

\subsubsection{Cooling limit as a function of frequency} 
\begin{figure}[ht]
\centering
\includegraphics[width=0.5\textwidth]{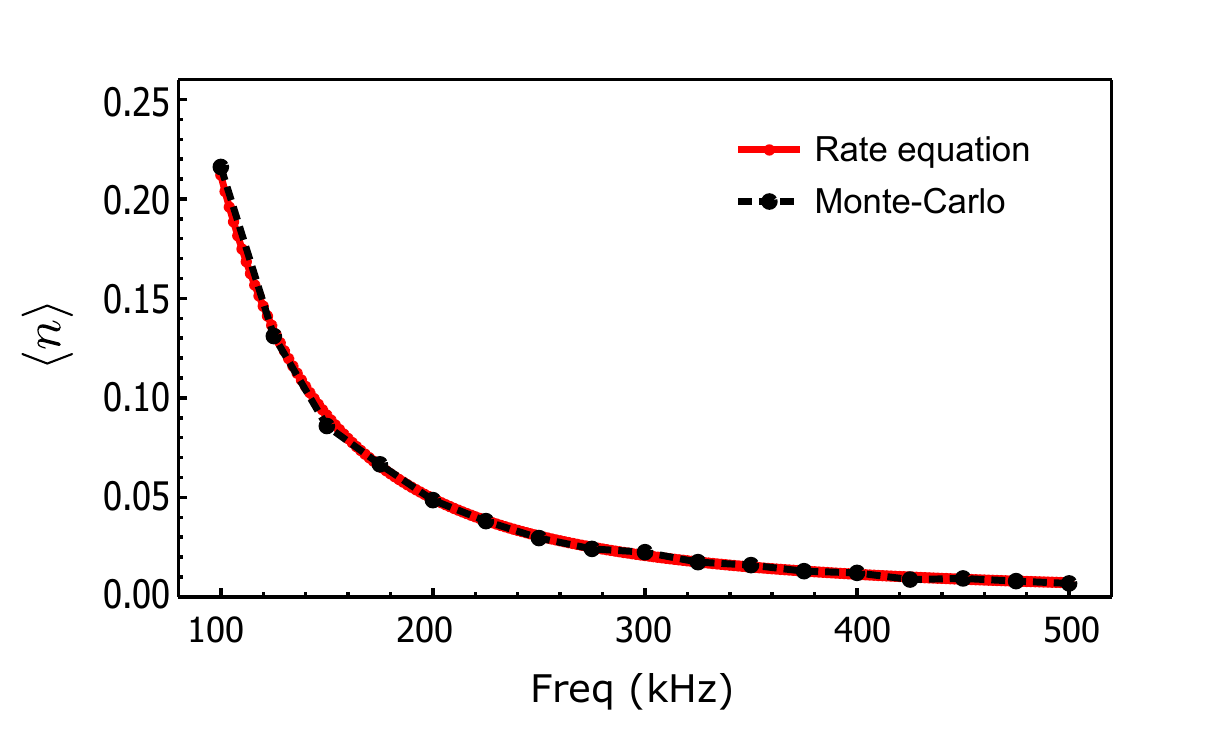}
\caption{Mean final occupation number as a function of ion oscillation frequency. The solid red curve corresponds to the cooling limit calculated using the rate equation approach, and the black points are the simulated cooling limits calculated from the Monte-Carlo approach.}
\label{nbarFreq}
\end{figure}

We also simulate the cooling limit as a function of motional frequency, shown in figure \ref{nbarFreq}. These simulations are carried out at $\Omega_0 =\tilde{\Gamma}= 2 \pi \times 40$ kHz. It can be seen that the mean phonon number increases rapidly towards lower motional frequencies. This can be attributed to increased diffusion as $\eta$ increases. The cooling limit depends upon the repumping rate and the rate at which 729 nm absorption takes place. To avoid off-resonant excitation, which can affect the cooling limit, the 729 nm/854 nm power can be reduced for the last period of sideband cooling. In general the cooling limit is found to be approximately inversely proportional to the square of the motional frequency at a fixed value of repumping and 729-nm absorption rates. 

The cooling limit for a single ion as a function of trap frequency is obtained via both the numerical solution of the rate equation and by simulating the equilibrium dynamics via a Monte-Carlo approach. The results are shown in figure 8, where the red solid line is the solution from rate equations and the black circles are the simulated cooling limit obtained via the Monte-Carlo method. Our studies show that both approaches give excellent agreement, which confirms the validity of the Monte-Carlo approach. In reality, for a single ion, both methods are competitive enough to carry out the simulations. However, for a multi-ion case, the rate equation approach is computationally expensive thus only the Monte-Carlo approach is viable. 

\subsection{Sideband heating: preparation of an ion outside the L-D regime} 
\begin{figure}[ht]
\centering
\includegraphics[width=0.45\textwidth]{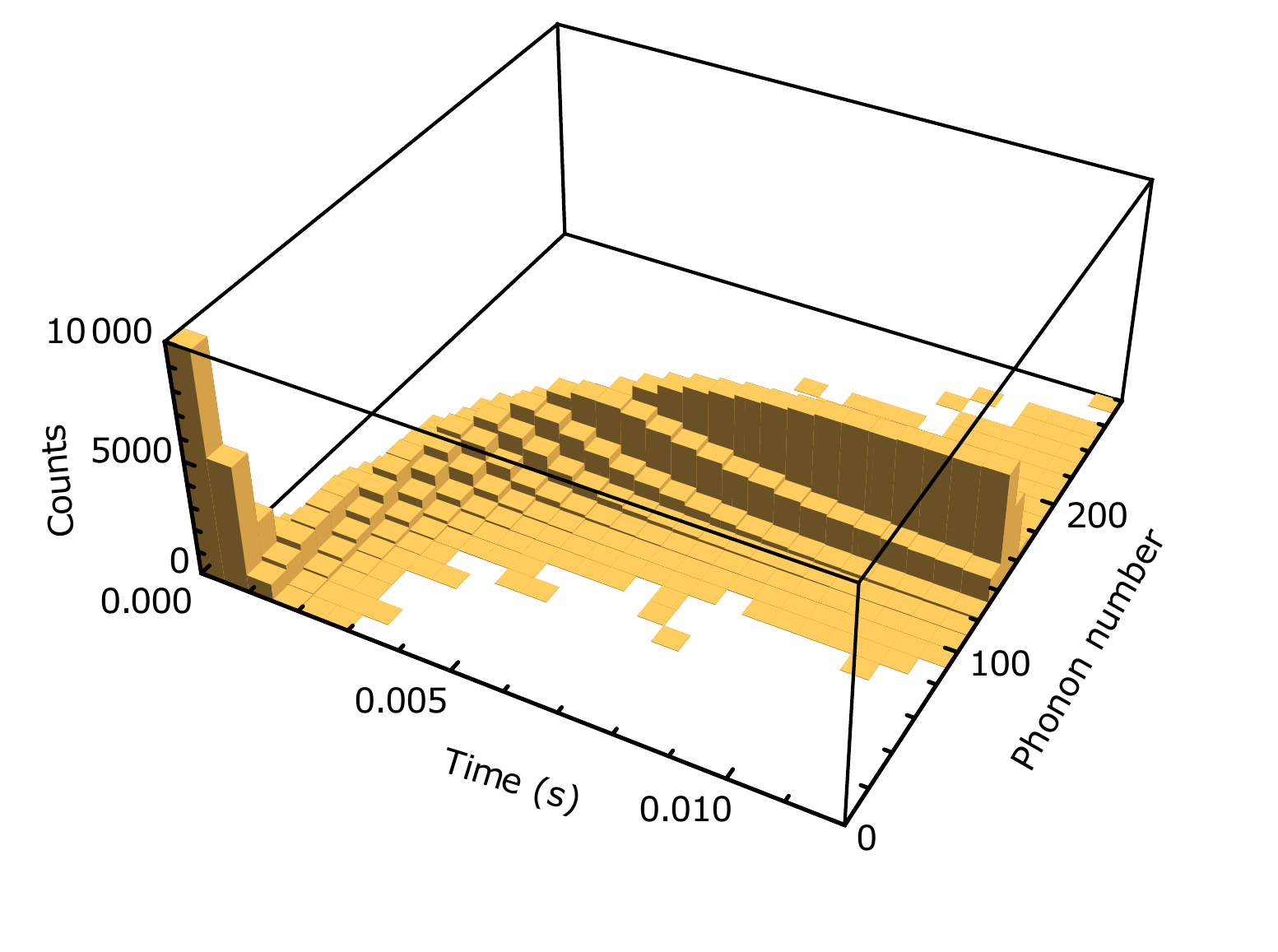}
\caption{A histogram plot of the population during sideband heating on the first blue sideband. It can be seen that the population moves towards the first minimum of the blue sideband (i.e. $n$ = 163 at $420$ kHz).}
\label{SBHHist}
\end{figure}

It can be interesting to prepare an ion in a high phonon state because higher order sidebands get stronger for high phonon numbers. In order to prepare an ion outside the L-D regime we tune the 729 nm laser to a blue sideband instead of a red sideband. To simulate that, we use a Monte Carlo approach and start from a ground state cooled ion population at a motional frequency of $420$ kHz. We then find the occupation probabilities of each Fock state as a function of time after starting the laser-ion interaction. In figure \ref{SBHHist}, we show the time evolution of the population during the sideband heating process. 

For further explanation of the sideband heating process, we extract the mean phonon number $\braket{n}$ and spread (standard deviation) as a function of sideband heating time. The values are indicated in figure 10. Initially, as the mean phonon number begins to increase with time, the standard deviation also begins to grow due to diffusion from spontaneous decay. Once the upper tail of the distribution begins to approach the minimum of the coupling strength at $n=163$, the standard deviation starts to shrink and the growth of the mean slows down due to the falling blue sideband coupling strength. Eventually, the mean settles to around $n=150$ and further heating only slowly increases the standard deviation as the population starts to leak through the sideband minimum to higher Fock states.

\begin{figure}[ht]
\centering
\includegraphics[width=0.45\textwidth]{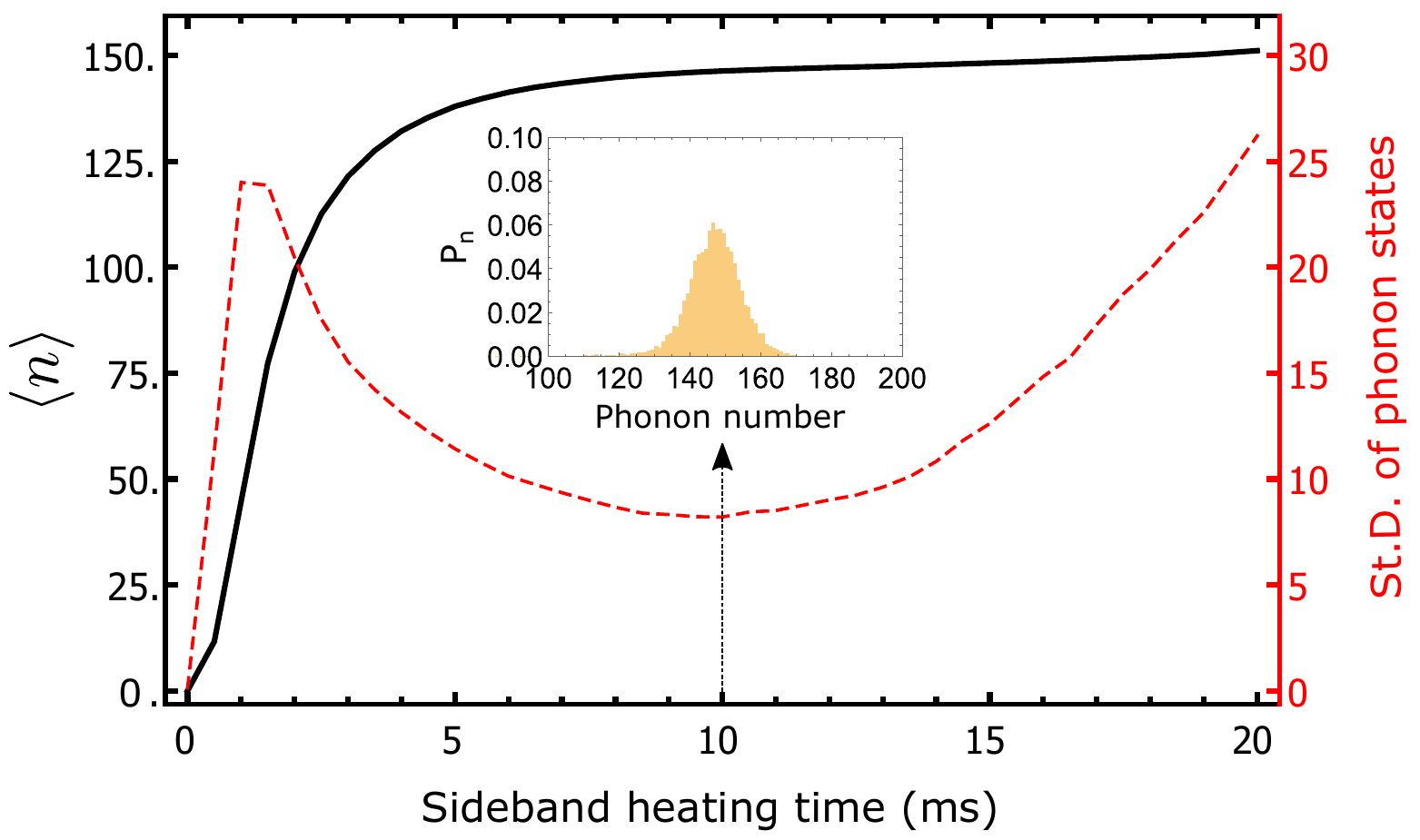}
\caption{Population dynamics after sideband heating as a function of laser interaction time. The solid black curve represents mean phonon number ($\braket{n}$) and the dashed red curve represents the standard deviation around the mean value. In the inset, population distribution after 10 ms of sideband heating. }
\label{SBHMeanPh}
\end{figure}

\begin{figure}[ht]
\centering
\includegraphics[width=0.45\textwidth]{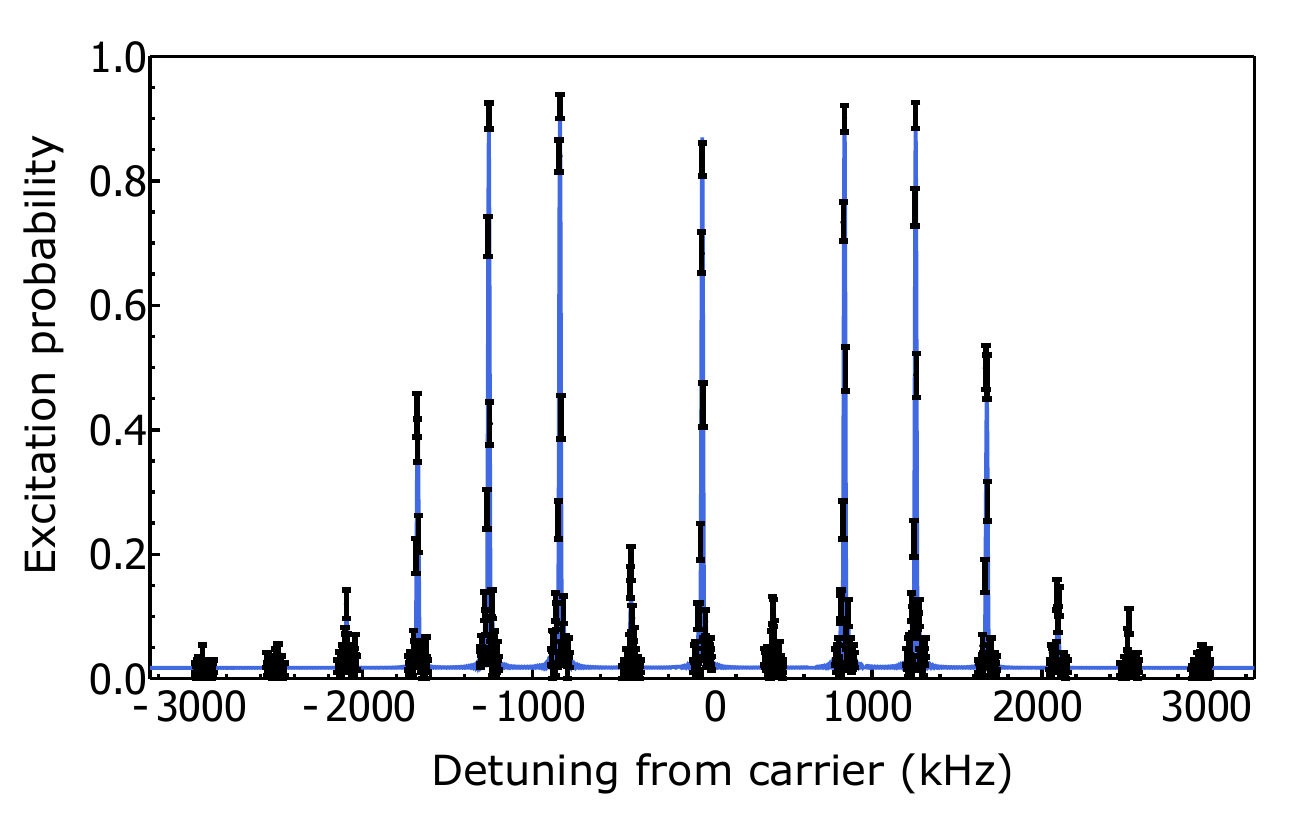}
\caption{Spectrum taken with a 65 $\mu$s probe time and Rabi frequency $\Omega_0=2\pi\times 16.5$ kHz after sideband heating for 10 ms, showing an ion prepared near to the first minimum of the blue sideband at $n= 163$. Note the near absence of the first order red and blue sidebands.}
\label{SBHExp}
\end{figure}

In order to verify this behavior we perform an experiment with a ground state cooled trapped $^{40}$Ca$^+$ ion in a Penning trap (see \cite{Stutter_2017} for a description of the experimental system). For optimized sideband heating laser parameters a spectrum is recorded and can be seen in figure \ref{SBHExp}. The experiment is performed at a motional frequency of $420$ kHz and the laser parameters are set to be the same as those used in the simulation.  As expected, the excitation for the first blue and red sidebands is seen to be much lower than that of the other sidebands. We also fitted the spectrum (shown in solid blue curve) and found  a mean phonon number $\braket{n} = 145 \pm 2$ and a spread in the population around the mean phonon number $\Delta n = 17$. The fitting is carried out using a Gaussian probability distribution function for the phonon occupations. The fitting model is independently validated through simulation studies. The experimental findings are in good agreement with the simulation.  

\subsection{Sideband cooling of a two-ion Coulomb crystal} Sideband cooling of multiple ions is significantly harder than that of a single ion due to the presence of multiple modes of oscillation which need to be cooled simultaneously. In this section we will study cooling dynamics of a two-ion crystal through a 2D Monte Carlo simulation where excitation on a  sideband of one motion will be associated with a carrier excitation on the other motion. 

\subsubsection{Two-ion chain} 
A two-ion chain in a Penning trap can only be obtained at a low axial trapping frequency. Increasing the axial frequency above a certain value flips the crystal to a planar orientation. We already know from the single-ion case that at a low axial frequency the system lies outside the L-D regime, hence we expect to observe population trapping during sideband cooling of the chain. In this case we can also avoid the population trapping by cooling on higher order sidebands. 

\begin{figure}[ht]
\centering
\includegraphics[width=0.5\textwidth]{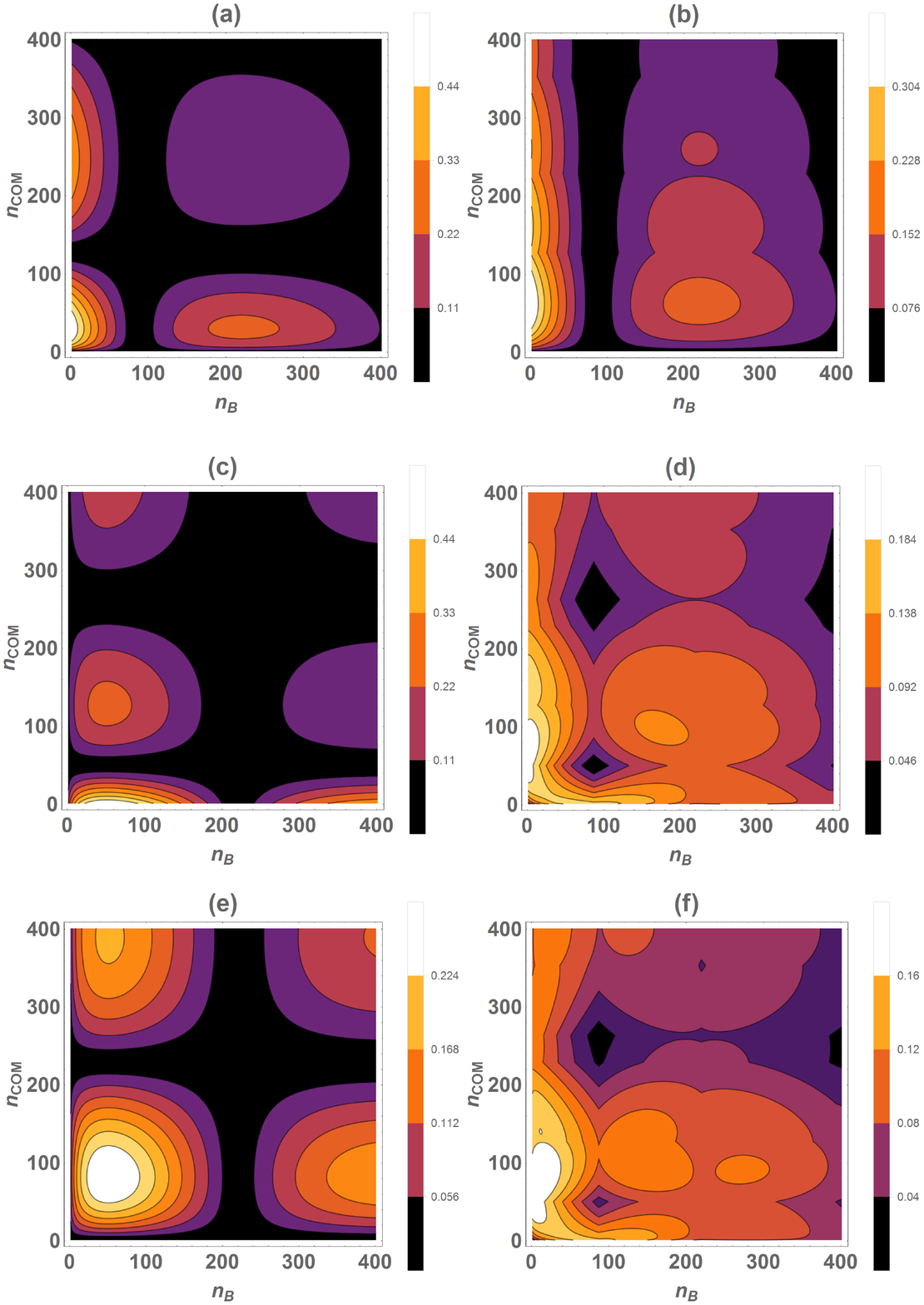}
\caption{Strength of various sidebands as a function of phonon number for a two-ion string at 162 kHz COM mode frequency. Starting from top, sideband strengths of (a) first red sideband of COM, (b) sum of first three COM red sidebands, (c) first breathing red sideband, (d) sum of first three COM and two breathing red sidebands, (e) first red breathing sideband of second red COM sideband and (f) sum of (e) and (d). Parts of this figure are reproduced from our earlier work (Journal of Modern  65:5-6, 549-559 (2018), reference \cite{Stutter_2017}) with permission from  Taylor \& Francis Ltd (\href{www.tandfonline.com}{www.tandfonline.com}).}
\label{ContPlot}
\end{figure}

To understand the cooling dynamics of a two ion chain we show contour plots of sideband strengths as a function of Fock state number at $\omega_{z} = 2 \pi \times162$ kHz (which gives $\omega_\text{B}=\sqrt{3} \omega_z=2 \pi \times 281$ kHz). In this particular case where the L-D parameter is fairly large ($\eta_\text{COM} = 0.17$ and $\eta_\text{B} = 0.13$), we expect to have a large fraction of the population beyond the first minimum of the red sideband for both of the motional modes \cite{Stutter_2017}. In order to bring all the population to the motional ground state, one has to target higher order sidebands repeatedly. A combined sideband contour plot of the first, second and third red sidebands of the COM mode along with the first red sideband of the breathing mode are shown in figure \ref{ContPlot} (a--c). Due to a smaller value of the L-D parameter for the breathing mode (i.e. $1/3^{1/4}$ times smaller than the COM mode), we expect to have less difficulty for the breathing mode than the COM  mode. We therefore find that the population can be taken to the ground state by targeting the first and second order breathing mode sidebands, whereas for the COM case we need the first, second and third order sidebands. 

The combined effect of all these five sidebands is shown in figure \ref{ContPlot} (d). Note that the combined plot still has darker areas, which correspond to the carrier excitation being a minimum for both of the motional modes.  This will lead to population trapping. In order to avoid population trapping in such regions one needs to target inter-modulation sidebands. A plot showing one of these intermodulation sidebands (i.e.\ the 2$^\text{nd}$ COM red sideband of the first breathing mode red sideband) is shown in figure \ref{ContPlot} (e). The overlap of all the sidebands mentioned above is finally shown in figure \ref{ContPlot} (f). It is clear that the above mentioned sequence will bring all the population to the ground state.

\begin{figure}[ht]
\centering
\includegraphics[width=0.50\textwidth]{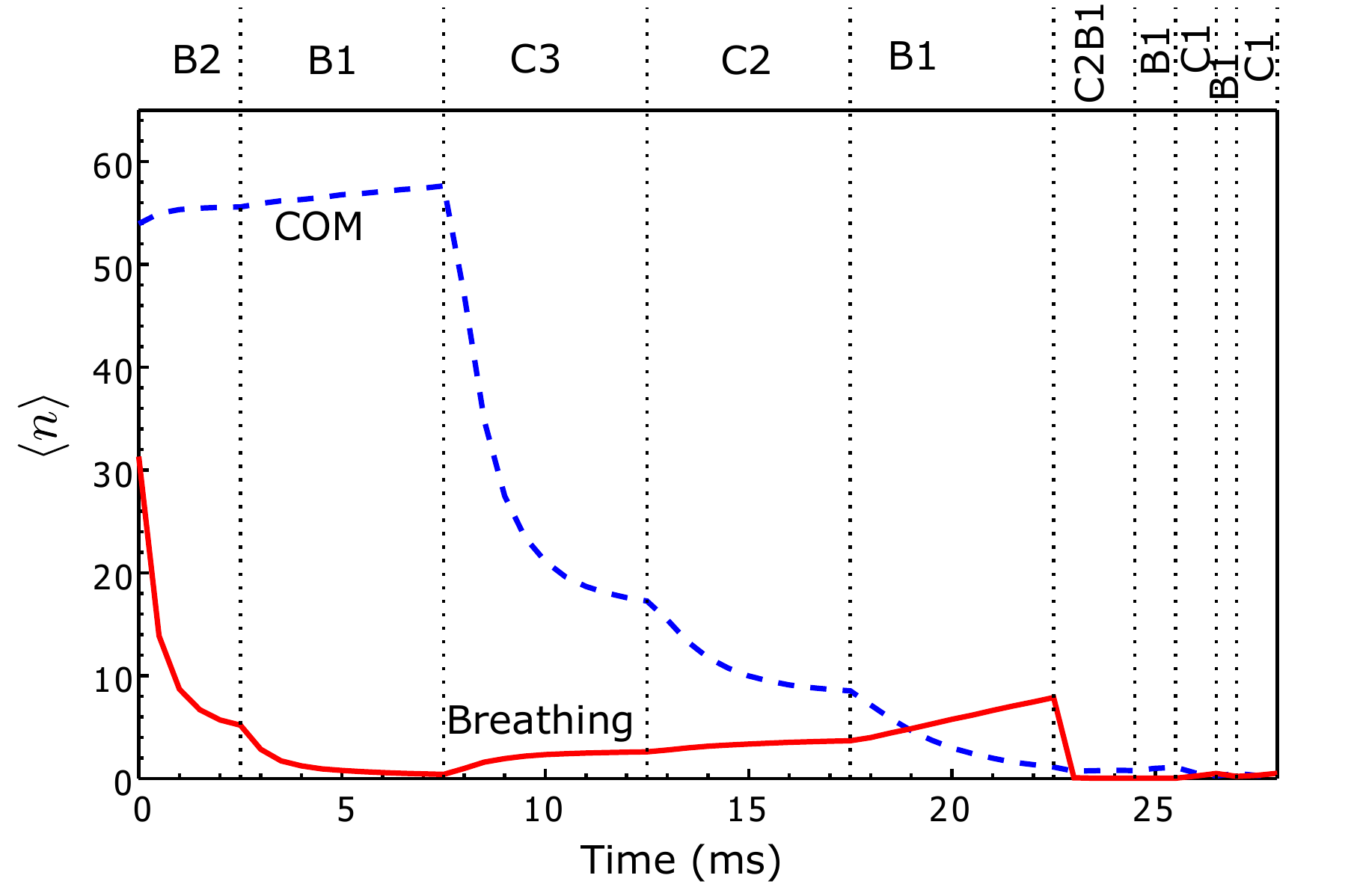} 
\caption{Mean phonon number of the COM mode (dashed blue) and the breathing mode (solid red) as a function of time during  multi-stage cooling of a two-ion chain with all sidebands included.  Label B/C denotes breathing/COM mode and indices show the red sideband used at each cooling stage.}
\label{SBCChain}
\end{figure}

Apart from population trapping, in the two-ion case we also observe heating on one mode while carrying out sideband cooling on the other  mode. This is mainly caused by the diffusion of population during the repumping process. {The cooling dynamics of a two ion chain as a function of cooling  time is shown in figure \ref{SBCChain}.} Intuitively, we start cooling on the breathing mode first, as it has a higher motional frequency (i.e. smaller L-D parameter). Hence we expect less heating for the breathing mode when we  switch cooling to the COM mode after achieving ground state cooling of the breathing mode. 

In order to cool the breathing mode to the ground state we start from the second order sideband and then switch to the first order sideband. Then we switch the cooling laser frequency to the third order sideband of the COM mode and progressively switch to second and first order sidebands. During that time we see heating of the breathing mode, hence we need to re-cool it. In order to clear out the population from the phonon states for which cooling for both motional modes ceases  (the common dark portion of the plot in figure \ref{ContPlot}(d)), an inter-modulation sideband is used in the cooling sequence. While these intermodulation sidebands are useful to prevent population trapping and cool both modes simultaneously, they are too weak to be used exclusively. The cooling sequence used in figure \ref{SBCChain} is devised from the qualitative discussion above; other sequences are also possible, giving similar results. These studies are consistent with our experimental results reported in reference \cite{Stutter_2017}. 

\subsubsection{Two-ion planar crystal} 
Sideband cooling of a planar crystal is of particular interest due to its possible use in quantum simulation \cite{Bohnet_2016, Kim_2010}. In the two-ion planar case the motional frequency of the out-of-phase mode (tilt mode) can be varied, as shown in equation \ref{freqT}, by changing the frequency of rotation of the crystal. This can be changed by locking the crystal to a rotating wall drive \cite{Huang_1998, Bharadia_2012} or by changing the perpendicular laser beam parameters (such as saturation, beam offset and detuning) and axialization field \cite{Bharadia_2012, Asprusten_2014}. The rotational frequency in a Penning trap can be varied between the magnetron frequency and the modified cyclotron frequency, and the minimum separation of the ions is obtained at $\omega_c/2$. 

\begin{figure}
\begin{centering}
\includegraphics[width=0.2\textwidth]{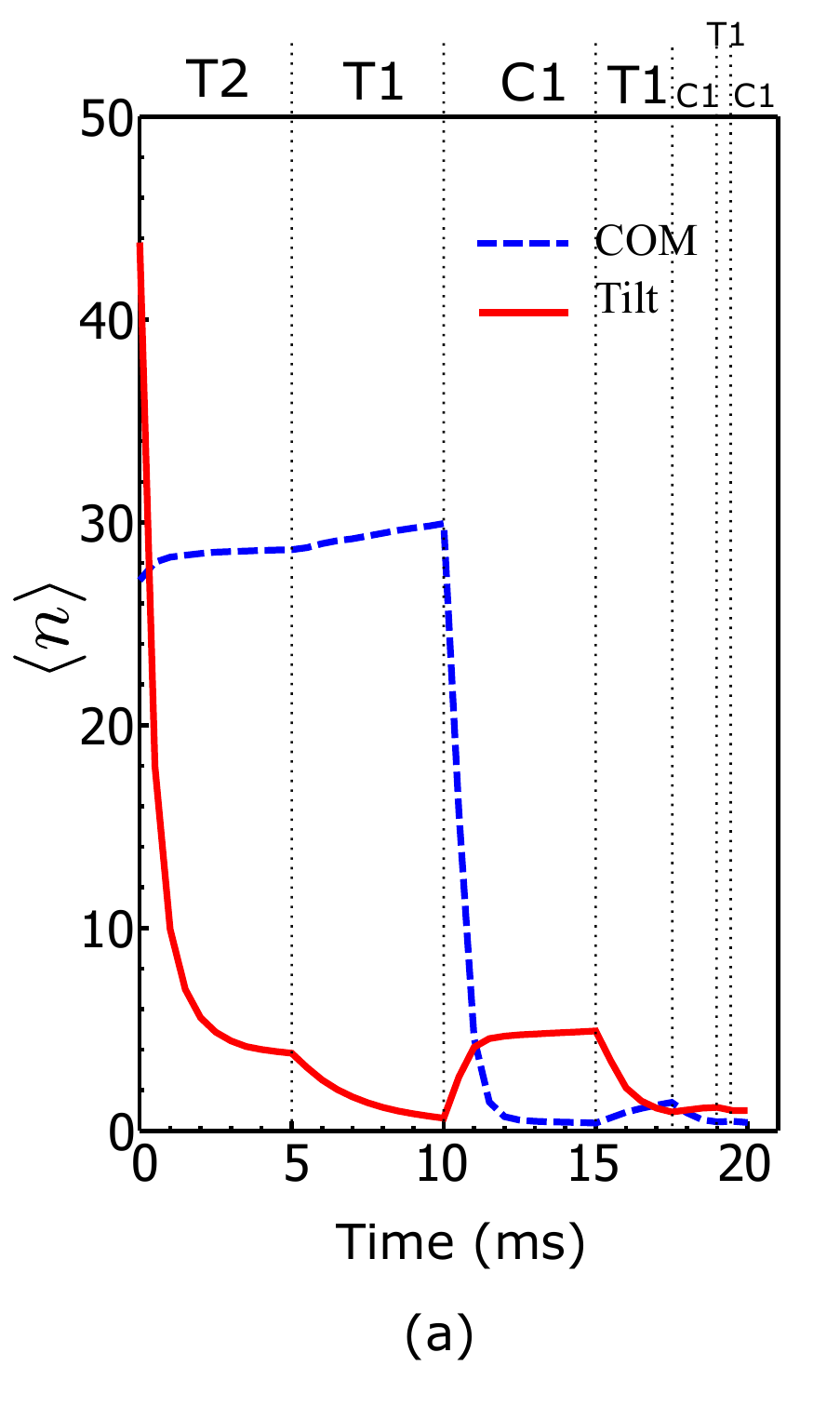}
\includegraphics[width=0.2\textwidth]{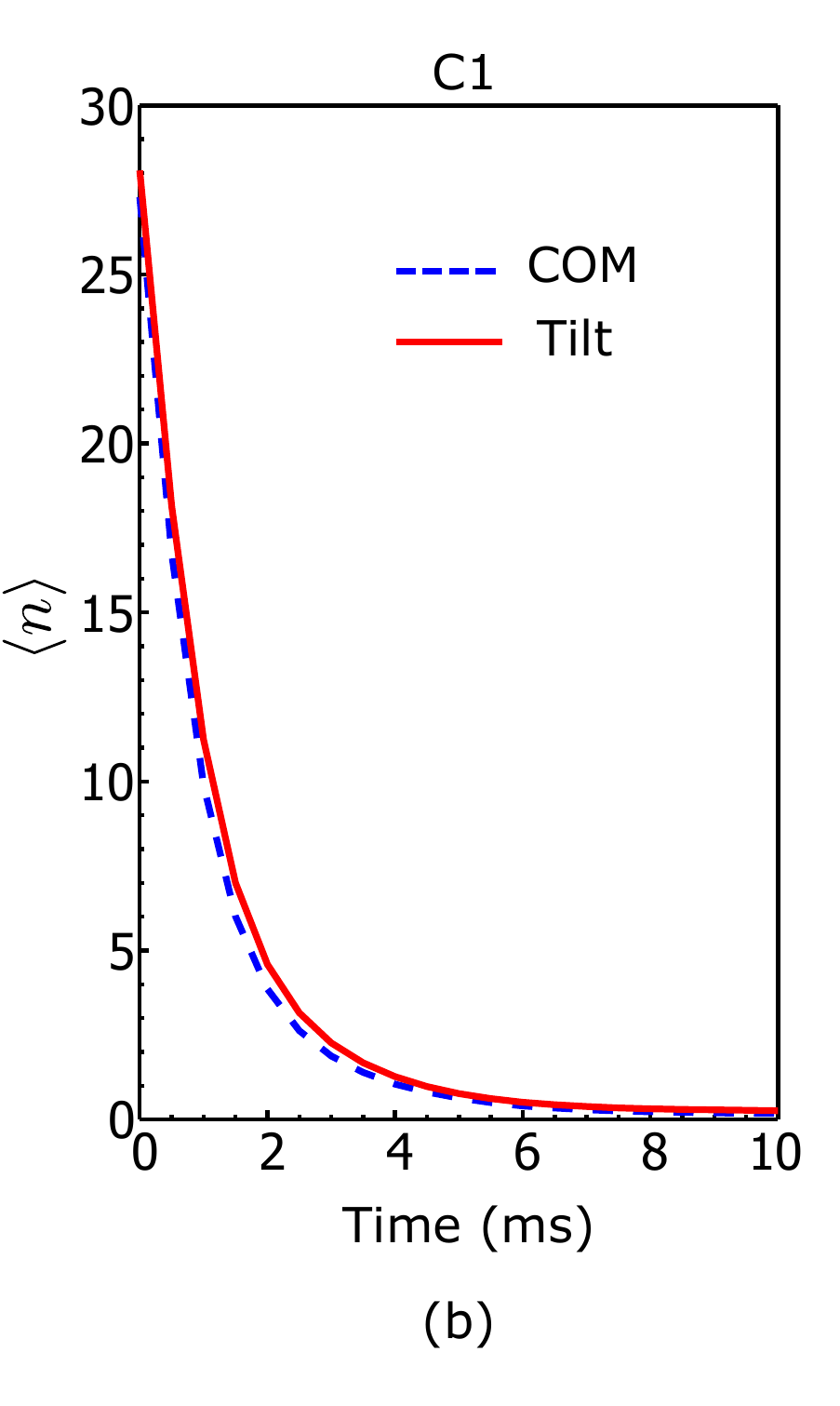}
\end{centering}
\caption{Mean phonon number as a function of sideband cooling time when COM and tilt motional modes are (a) resolved and (b) unresolved. Note that off-resonant cooling takes place for the tilt mode when the laser is tuned to the COM mode. Label T/C denotes tilt/COM mode and indices show the red sideband used at each cooling stage. } \label{CoolingPlanar}
\end{figure}

At first we simulate the cooling dynamics at fixed laser cooling parameters for a two-ion planar crystal in the resolved motional modes case. Laser parameters for this case are chosen to be the same as for the two-ion chain. The axial COM mode frequency is chosen to be $346$ kHz, where the  tilt mode frequency can have values from $216$ kHz to $346$ kHz, depending upon the rotation frequency of the crystal. In the resolved modes case we assume the tilt mode is at $216$ kHz. This condition can be achieved at a crystal rotation frequency equal to $\omega_{c}/2$. In this case the L-D parameters are $\eta_\text{COM} = 0.116$ and $\eta_\text{tilt} = 0.146$. The simulated cooling dynamics is shown in figure \ref{CoolingPlanar} (a). It can be seen that both modes can be ground state cooled by using the second and first red sidebands on the tilt mode followed by first on the COM mode.  In this case we also see sideband heating on the COM mode while attempting to cool the breathing mode and vice versa. At the end  we also repeat the lower order sidebands in order to efficiently cool both modes to the ground state. In this case, the intermodulation sidebands are not important due to higher trapping frequency and an insignificant amount of population trapping, hence cooling on the intermodulation sideband is not considered in the simulations. 

In the other case we choose the motional frequency of tilt motional mode to be close to the COM mode,  by tuning the rotation frequency slightly above the magnetron frequency. For the simulation we chose the tilt mode frequency to be $339$ kHz and COM mode frequency to be $346$ kHz and the  L-D parameters are then $\eta_{COM} = 0.116$ and $\eta_{tilt} = 0.118$. This is similar to what we see in experiments where we are unable to apply a suitable rotating wall to lock the crystal rotation frequency. In this case the sidebands are not expected to be well resolved after Doppler cooling, hence cooling on one of the motional modes can simultaneously cool the other mode off-resonantly. A simulated cooling plot is shown is figure \ref{CoolingPlanar} (b). It can be seen that  by setting  the cooling laser frequency to the COM mode we obtain simultaneous ground state cooling of both modes. The experimental results for this case can be seen in reference \cite{Stutter_2017}.

\section{Conclusion} In this article we presented simulation studies of cooling dynamics of trapped ions outside the L-D regime. Furthermore, we extended the treatment to include a realistic five-level system of a $^{40}$Ca$^+$ ion in a Penning trap. To start with, we optimized the cooling parameters by simulating cooling dynamics of a single ion. We also showed the importance of population trapping as a consequence of starting outside the L-D regime and how to overcome it by using composite cooling sequences involving higher order sidebands. We also simulated sideband heating, where an ion was heated near to a Fock state number where the sideband strength approaches zero. One potential application is  to study the quantum behavior of trapped ions far outside the L-D regime. See reference \cite{Stutter_2017} for related experimental demonstrations.

We also extended the simulation studies to multiple ions. Two configurations of two-ion Coulomb crystals were studied. The studies included the case of unresolved sidebands where the tilt mode frequency is close to the COM frequency. Hence, fewer sidebands needed to be targeted in order to achieve ground state cooling.    
    
Studies of sideband cooling outside the L-D regime might be useful for sympathetic ground state cooling of other species (e.g. protons or heavy atomic or molecular ions), where the motional frequency of a combined motional mode could be small. In such a system the L-D parameter may have a large value. The simulations reported here show that the motional ground state can be achieved for two-ion crystals in a Penning trap, even when the system starts well outside the L-D regime. Our study of sideband cooling outside the L-D regime also takes into account multiple atomic levels, and we have shown that it is still possible to achieve ground state cooling. Penning traps have already been found to have low anomalous heating rates and the feasibility of sideband cooling for small Coulomb crystals opens paths to use them for precision spectroscopy and high fidelity quantum simulators.

\section{Acknowledgments} We would like to acknowledge Prof.\ Danny Segal's inspirational and innovative guidance in the development of these trapped ion studies at Imperial College London. The research leading to these results has received funding from the People Programme (Marie Curie Actions) of the European Union's Seventh Framework Programme (FP7/2007-2013) under REA grant agreement no 31723. This work was supported by the UK Engineering and Physical Sciences Research Council (Grant EP/L016524/1). 

\bibliography{BBLAPSpaper}

\end{document}